\documentclass[10pt,letterpaper]{article}
\usepackage[margin=1in]{geometry}
\usepackage{ifpdf}
\ifpdf
    \usepackage[pdftex]{graphicx}
    \graphicspath{{figs/}{matlab/}}   
    \usepackage[update]{epstopdf}
\else
	\usepackage{graphicx}
	\graphicspath{{figs/}{matlab/}}   
\fi
\usepackage{pdflscape}
\usepackage{marvosym}
\usepackage{enumitem,color}
\usepackage{amsthm}

\usepackage{hyperref}
\usepackage{array}
\usepackage{amsmath,amsthm}
\usepackage{amsfonts}
\usepackage{amssymb}
\usepackage{cite,setspace}
\allowdisplaybreaks
\usepackage{multirow}
\newcommand{\code}[1]{\texttt{#1}}
\newcommand\tab[1][0.75cm]{\hspace*{#1}}

\begin{document}

\title{User Manual CAI version-1.0: An Open-Source Toolbox for Computer-Aided Investigation on the Fundamental Limits of Information Systems}
\author{Chao Tian, James Plank, Brent Hurst, and Ruida Zhou}
\maketitle

\vspace{3cm}

Summary of main changes and additions in version 1.0 of the CAI toolbox:
\begin{enumerate}
\item A new and more powerful parser replaces the older parser in this version, based on the popular JSON format. This change allows more efficient processing and representation of the problem description, which significantly reduces the processing time before the main computation engine is invoked. Utilities to convert between the new format and older format are provided in the toolbox.  
\item A new function is added which allows the user to set up the problem such that the values of various information measures can be extracted for the computed optimal LP solution.
\item A new function is added which allows the user to set up the problem such that lower bound and upper bound of specified information measures are computed, for which the optimal value of the objective function does not change (i.e., a sensitivity analysis). 
\item An improved data structure is used in the core symmetry and functional reduction step, which improves the performance of the original algorithm significantly.
\item A Python template is included for a complete processing pipeline, from problem description generation to tradeoff regime visualization, of several well known coding problems in information theory and coding theory: erasure coding, coded caching, private information retrieval, and secret sharing.  
\end{enumerate}
This manual has been updated accordingly to reflect these changes. In addition, in contrast to the previous version, where only a small set of relevant prior works were discussed, a more complete literature review is included in this update. 

\newpage

\begin{abstract}
We provide an open source toolbox on \url{https://github.com/ct2641/CAI/releases/tag/1.0} to conduct computer-aided investigation on the fundamental limits of information systems. The toolbox relies on either Gurobi or Cplex as the linear program solving engine. The program can read a problem description file, and then fulfill the following tasks: 1) compute a bound for a given linear combination of information measures and provide the value of information measures at the optimal solution; 2) efficiently compute a polytope tradeoff outer bound between two information quantities; 3) produce a proof (as a weighted sum of known information inequalities; and 4) provide the range for information quantities between which the optimal value does not change (sensitivity analysis). This technical report provides an overview of this toolbox, a detailed description of the syntax of the problem description file, and a few example use cases. 
\end{abstract}

\section{Introduction}
One of the most distinguishing features of information theory is its ability to provide fundamental limits to various communication and computation systems, which may be extremely difficult, if not impossible, to establish otherwise. There are a set of well-known information inequalities, such as the non-negativity of mutual information and conditional mutual information, which are guaranteed to hold simply due to the basic mathematical properties of the information measures such as entropy and conditional mutual information. Fundamental limits of various information systems can be obtained by combining these inequalities strategically. The universality of the information measures implies that fundamental limits of diverse information systems can be derived in a general manner. 

Conventionally, the proofs for such fundamental limits are hand-crafted and written as a chain of inequalities, where each individual step is one of the afore-mentioned known information inequalities, or certain equality and inequalities implied by the specific problem settings. As information systems become more and more complex, such manual efforts have become increasingly unwieldy, and computer-aided approaches naturally emerge as possible alternatives. A computer-aided approach can be particularly attractive and productive during the stage of initial problem exploration and when the complexity of the system prevents an effective bound to be constructed manually. 

The purpose of the toolbox is to help researchers in the field better utilize this computer-aided approach in their own research, and also to foster the discovery of more advanced computer-aided techniques. This toolbox is designed to provide a streamlined process of several already relatively mature methods. Generally speaking, if a problem can be represented by relations among the information measures, then the program can read a problem description text file and produce the desired bounds, certain analysis details, tradeoff region, and proofs, assuming the scale of the problem does not exceed certain memory and word length restrictions. It is our hope that by opening up the source code, researchers will be able to find more creative techniques to further advance this computer-aided approach. 

\section{Literature Review}

In a pioneer work, Yeung \cite{Yeung:97} pointed out and demonstrated how a linear programming (LP) framework can be used to computationally verify whether an information inequality involving Shannon's information measures is true not not, or more precisely, whether it can be proved using a general (yet not complete) set of known information inequalities, which is referred to as Shannon-type inequalities. A Matlab implementation based on this connection, called the information theory inequality prover (ITIP) \cite{ITIP}, was made available online. A subsequent effort by another group (XITIP \cite{XITIP}) replaced the Matlab LP solver with a more efficient open source LP solver and also introduced a more user-friendly interface. Later on,  a new version of ITIP also adopted a more efficient LP solver to improve the computation efficiency. ITIP and XITIP  played important roles in the study of non-Shannon-type inequalities and Markov random fields \cite{yeung2002information, dougherty2006six,dougherty2011non}. 

Despite its significant impact, ITIP is a generic inequality prover, and utilizing it on any specific coding problem can be a daunting task.  It can also fail to provide meaningful results due to the associated computation cost. Instead of using the LP to verify a hypothesized inequality, a more desirable approach is to directly use a computational approach on the specific problem of interest to find the fundamental limits, and moreover, to utilize the inherent problem structure in reducing the computation burden. This was the approach taken on several problems of recent interest, such as distributed storage, coded caching and private information retrieval, by one of the authors \cite{Tian:JSAC13,TianLiu:15,Tian:16Computer,tian2018caching,Tian_Sun_Chen_Storage,tian2020storage}, and it was shown to be rather effective. 

One key difference in the above-mentioned line of work, compared to several other efforts in the literature, is the following. Since most information theoretic problem of practical relevance or current interests would induce a quite large LP instance, considerable effort was given to reducing the number of LP variables and the number of LP constraints algorithmically, before the LP solver is even invoked. Particularly, symmetry and problem-specific dependence were used explicitly for this purpose, instead of the standard approach of leaving them for the LP solver to eliminate. This approach allows the program to handle larger problems than ITIP could, which yielded meaningful results on problems of current interests. Moreover, through LP duality, it was demonstrated in \cite{Tian:JSAC13} that human-readable proofs can be generated by taking advantage of the dual LP. This approach of generating proofs was adopted and extended by several other works \cite{li2016multilevel,ho2020proving}.

From more theoretical perspectives, a minimum set of LP constraints under problem-specific dependence was fully characterized in \cite{chan2019minimal}, and the problem of counting the number of LP variables and constraints after applying problem specific symmetry relations was considered in \cite{ZhangTian:17TCOM}. However, these results do not lead to algorithmic advantage, since the former relies on a set of relationship testings which is algorithmically expensive, and the latter 
 provided a method of counting instead of enumerating these information inequalities.

Li et al.~used a similar computational approach to tackle the multilevel diversity coding problem \cite{Walsh:16} and multi-source network coding problems with simple network topology \cite{li2017multi} (see also \cite{apte2015exploiting}); however the main focus was to provide efficient enumeration and classification of the large number of specific small cases (all cases considered require 7 or fewer random variables) where each case itself does not have a symmetry structure. In contrast, the focus of the toolbox in our work is on larger problem settings (involving 16-25 random variables) where significant symmetry is present in the coding problem itself, although problems without symmetry can also be handled. Beyond computing outer bounds, the problem of computationally generating inner bounds was also explored \cite{li2013new,apte2014algorithms}. 

Recently, Ho et al. \cite{ho2020proving} revisited the problem of using the LP framework for verifying the validity of information inequalities, and proposed a method to computationally disprove certain information inequality. Moreover, it was shown that the alternating direction method of multipliers (ADMM) can be used to speed up the LP computation. In a different application of the LP framework \cite{Permuter:15newsletter},  Gattegno et al. used it to improve the efficiency of the Fourier-Motzkin elimination procedure often encountered in information theoretic study of multiterminal coding problems. In another generalization of the approach, Gurpinar and Romashchenko used the computational approach in an extended probability space such that information inequalities beyond Shannon-types may become active \cite{gurpinar2019use}. 

\section{Entropy, Entropy LP Formulation, and Reductions}

Although it is not strictly required to understand the mechanism behind the computer-aided approach to use this toolbox, such knowledge may become critical when a more efficient problem formulation is required or additional customized functionalities are needed in the toolbox. It also helps the readers understand the various keys in the problem description file we will discuss later. For this reason, we include a brief overview in this section; readers are referred to \cite{CoverThomas,Yeung:book,yeung2008information} for more details on information measures and the entropy linear programming (LP) problem.

\subsection{Information Measures}
The entropy of a random variable $X$ distributed on a finite alphabet $\mathcal{X}$ is a mapping from a probability mass function to a real value. Denote the probability mass for any given letter $x\in\mathcal{X}$ as $p(x)$, then the entropy of the random variable $X$ is defined as 
\begin{align}
H(X)=-\sum_{x\in\mathcal{X}} p(x)\log p(x),
\end{align}
where we take the convention of defining $a\log a=0$ when $a=0$. Although we referred to $H(X)$ as the entropy of the random variable $X$, it is perhaps more accurate to refer to it as the entropy of a given probability mass function. The entropy of $X$ is usually understood as a measure of the expected uncertainty in the random variable $X$. The joint entropy of $n$ random variables $(X_1,X_2,\ldots,X_n)$ is defined similarly as
\begin{align}
H(X_1,X_2,\ldots,X_n)=-\sum_{(x_1,x_2,\ldots,x_n)\in\prod_{i=1}^n\mathcal{X}_n} p(x_1,x_2,\ldots,x_n)\log p(x_1,x_2,\ldots,x_n).
\end{align}
The same way of generalization can be used for the information measures in the discussion that follows, and thus we only provide the basic definitions but not the generalized ones next. 

The conditional entropy of two random variables $X_1$ given $X_2$ is defined as
\begin{align}
H(X_1|X_2)=H(X_1,X_2)-H(X_2),\label{eqn:conditional}
\end{align}
which is understood as a measure of the expected uncertainty in $X_1$ when $X_2$ is known. The mutual information between $X_1$ and $X_2$ is defined as
\begin{align}
I(X_1;X_2)=I(X_2;X_1)=H(X_1)+H(X_2)-H(X_1,X_2),
\end{align}
which is understood as the information in $X_1$ about $X_2$, and vice versa. Similarly, the conditional mutual information is given as
\begin{align}
I(X_1;X_2|X_3)&=H(X_1|X_3)+H(X_2|X_3)-H(X_1,X_2|X_3)\notag\\
&=H(X_1,X_3)+H(X_2,X_3)-H(X_1,X_2,X_3)-H(X_3).\label{eqn:conditionalMI}
\end{align}

For random variables with a continuous alphabet, the corresponding definitions are differential entropy, joint differential entropy, conditional differential entropy, and mutual information, however, in this work we will only work with random variables with finite and discrete alphabets. 

\subsection{Information Inequalities}

The most well-known information inequalities are based on the non-negativity of the conditional entropy and mutual information, which are
\begin{align}
&H(X_1|X_2)\geq 0\notag\\
&I(X_1;X_2|X_3)\geq0, \label{eqn:BaseIneq}
\end{align}
where the single random variables $X_1$, $X_2$, and $X_3$ can be replaced by sets of random variables. A very large number of inequalities can be written this way, when the problem involves a total of $n$ random variables $X_1,X_2,\ldots,X_n$. These inequalities are referred to as Shannon-type inequalities \cite{Yeung:97}. Within the set of all information inequalities in the form shown in (\ref{eqn:BaseIneq}), many are implied by others. There are also other information inequalities implied by the basic mathematical properties of the information measure but not in these forms or directly implied by them, which are usually referred to as non-Shannon-type inequalities. Non-Shannon-type inequalities are notoriously difficult to generate and utilize \cite{Zhang:97,matus2007infinitely,Dougherty:05,dougherty2007networks}. In practice, most bounds on the fundamental limits of information systems are derived using only Shannon-type inequalities, and the toolbox is designed around them. 

\subsection{The Entropy LP Formulation}

Suppose we express all the relevant quantities in a particular information system (a coding problem) as random variables $(X_1,X_2,\ldots,X_n)$, e.g., $X_1$ is an information source and $X_3$ is its encoded version at a given point in the system. 
In this case, the derivation of an fundamental limit in an information system or a communication system can conceptually be understood as the following optimization problem:
\begin{align*}
\text{minimize: }&\text{a weighted sum of certain joint entropies} \\
\text{subject to: }&\text{(I) generic constraints that any information measures must satisfy}\\
&\text{(II) problem specific constraints on the information measures},
\end{align*}
where the variables in this optimization problem are all the \textbf{information measures} on the random variables $X_1,X_2,\ldots,X_n$ that we can write down in this problem. For example, if $H(X_2,X_3)$ is certain quantity that we wish to minimize (e.g., as the total amount of the compressed information in the system), then the solution of the optimization problem with $H(X_2,X_3)$ being the objective function will provide the fundamental limit of this quantity (e.g., the lowest amount we can compress the information to). 

The first observation is that the variables in the optimization problem above only need to involve all joint entropies but not others, i.e., $2^n-1$ quantities in the form of $H(X_\mathcal{A})$ where $\mathcal{A}\subseteq \{1,2,\ldots,n\}$ and $X_{\mathcal{A}}=\{X_i,i\in\mathcal{A}\}$. The reason that we do not need to include conditional entropy, mutual information, or conditional mutual information in the problem is simple: they can be written simply as a linear combination of the joint entropies as given in (\ref{eqn:conditional})-(\ref{eqn:conditionalMI}).

Next let us focus on the two classes of constraints. To obtain a good (hopefully tight) bound, we wish to include all the Shannon-type-inequalities as generic constraints in the first group of constraints. However, enumerating all of them is not the best approach, as we have mentioned earlier that there are redundant inequalities that are implied by others. In a mathematical optimization problem, we wish to use a concisely represented constrained set, and an immediate question is which set is the most concise. Yeung identified such a minimal set of constraints which are called elemental inequalities \cite{Yeung:97,Yeung:book}:
\begin{align}
&H(X_i|X_{\mathcal{A}})\geq 0,\quad  i\in \{1,2,\ldots,n\},\qquad\mathcal{A}\subseteq \{1,2,\ldots,n\}\setminus \{i\}\label{eqn:Shannontype1}\\
&I(X_i;X_j|X_{\mathcal{A}})\geq 0, \quad i\neq j,\, i,j\in \{1,2,\ldots,n\},\qquad\mathcal{A}\subseteq \{1,2,\ldots,n\}\setminus\{i,j\}\label{eqn:Shannontype2}
\end{align}
Note that both (\ref{eqn:Shannontype1}) and (\ref{eqn:Shannontype2}) can be written as a linear constraint in terms of joint entropies. It is straightforward to see that there are $n+{n \choose 2} 2^{n-2}$ elemental inequalities. These are the generic constraints that we will use in group (I). 

The second group of constraints are the problem specific constraints. These are usually the implication relation required by the system or the specific coding requirements. For example, if $X_4$ is a coded representation of $X_1$ and $X_2$, then this relation can be represented as 
\begin{align}
H(X_4|X_1,X_2)=H(X_1,X_2,X_4)-H(X_1,X_2)=0,\label{eqn:implication}
\end{align}
which is a linear constraint. This group of constraints may also include independence and conditional independence relations. For example, if $X_1,X_3,X_7$ are three mutually independent sources, then this relation can be represented as
\begin{align}
H(X_1,X_3,X_7)-H(X_1)-H(X_3)-H(X_7)=0,
\end{align}
which is also a linear constraints. In the examples in later sections, we will provide these constraints more specifically. 

The two groups of constraints are both linear in terms of the optimization problem variables, i.e., the $2^n-1$ joint entropies (defined on the $n$ random variables), and thus we have a linear program (LP) at hand. This optimization problem forms the core behind the various computational functionalities of this toolbox. 

\section{Reduced Entropy LP}

In this section, we discuss two techniques to reduce the complexity of the entropy LP. Without using these techniques, many information system or coding problems of practical interest appear too complex to be solved directly in the entropy LP formulation. In order to be more specific, we first introduce two working examples that will be used throughout this document to illustrate the main idea. 

\subsection{Two Examples}

\begin{figure}[t]
\centering
\includegraphics[width=0.25\textwidth]{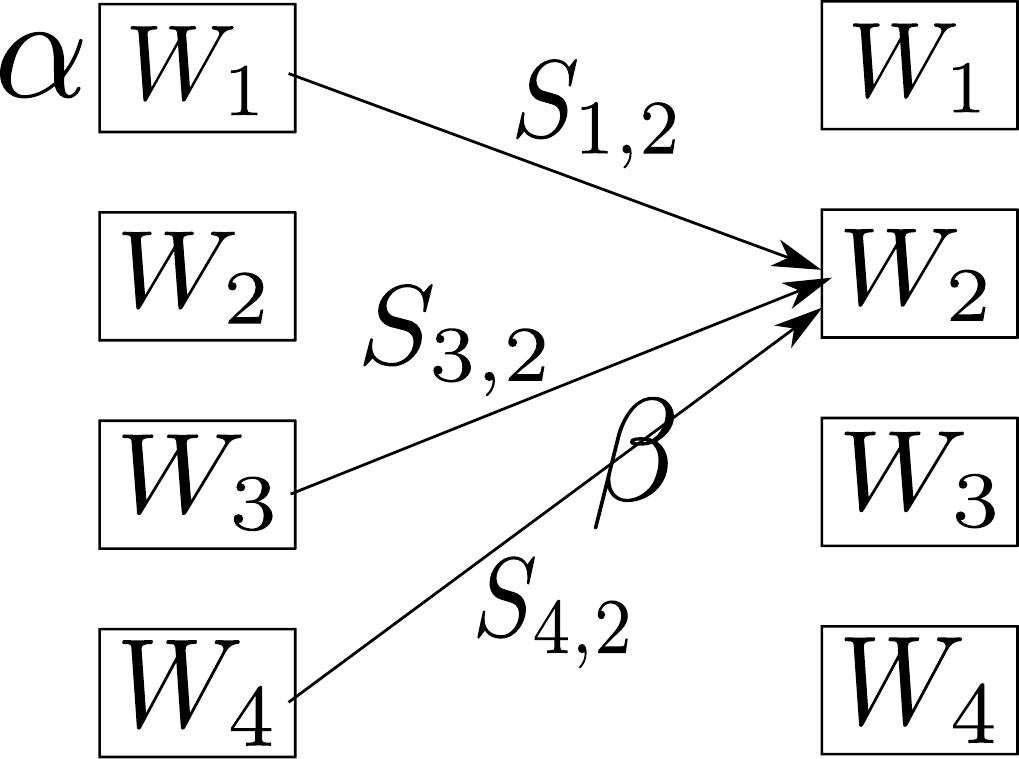}
\caption{The regenerating code problem with $(n,k,d)=(4,3,3)$. \label{fig:rg}}
\end{figure}

\begin{figure}[t]
\centering
\includegraphics[width=0.6\textwidth]{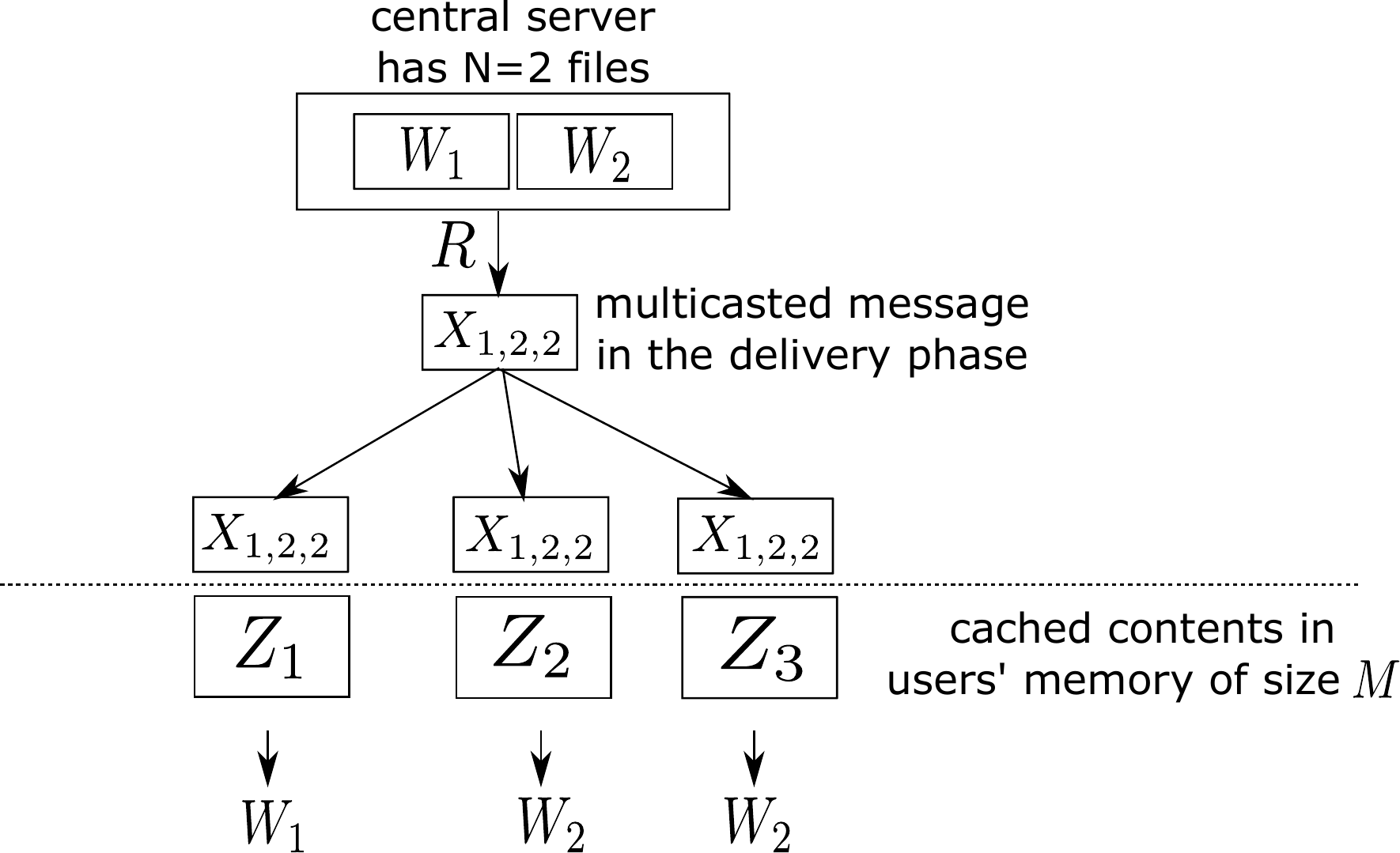}
\caption{The caching problem with $(N,K)=(2,3)$. \label{fig:caching}}
\end{figure}

The two problems we discuss next are the regenerating code problem and the coded caching problem: 
\begin{itemize}
\item The $(n,k,d)$ regenerating code problem \cite{dgw:10:nc,drw:11:snc} considers the situation a message is stored in a distributed manner in $n$ nodes, each having capacity $\alpha$. Two coding requirements need to be satisfied: 1) the message can be recovered from any $k$ nodes, and 2) any single node can be repaired by downloading $\beta$ amount of information from any $d$ remaining nodes each. The fundamental limit of interest is the optimal tradeoff between the storage cost $\alpha$ and the download cost $\beta$; see Fig. \ref{fig:rg}. We will use the $(n,k,d)=(4,3,3)$ case as our working example. In this setting, the stored contents as $W_1,W_2,W_3,W_4$, and the repair message sent from node $i$ to repair $j$ is denoted as $S_{i,j}$. In this case, the set of the random variables in the problem are
\begin{align*}
W_1,W_2,W_3,W_4,S_{1,2},S_{1,3},S_{1,4},S_{2,1},S_{2,3},S_{2,4},S_{3,1},S_{3,2},S_{3,4},S_{4,1},S_{4,2},S_{4,3}.
\end{align*}
Some readers may notice that we do not include in this setting a random variable to represent the original message stored in the system. This is because it can be equivalently viewed as the collection of $(W_1,W_2,W_3,W_4)$ and can thus be omitted in this formulation. Later on, we will discuss another layer of simplification when dealing with $(n,k,d)=(5,4,4)$. 

\item The $(N,K)$ coded caching problem \cite{MaddahAliNiesen:14} considers the situation that a server, which holds a total $N$ mutually independent files of unit size each, serves a set of $K$ users, each with a local cache size $M$. The users can prefetch some content, but when they reveal their requests, the server must multicast certain common information of size $R$. The requests are not revealed to the server beforehand, and the prefetching must be designed to handle all cases. The fundamental limit of interest is the optimal tradeoff between the cache capacity $M$ and the transmission size $R$; see Fig. \ref{fig:caching}. In this setting, the messages are denoted as $(W_1,W_2,\ldots,W_N)$, the prefetched contents as $(Z_1,Z_2,\ldots,Z_K)$, and the transmission when the users requests $(d_1,d_2,\ldots,d_K)$ is written as $X_{d_1,d_2,\ldots,d_K}$. We will use the case $(N,K)=(2,3)$  as our second running example in the sequel, and in this case the random variables in the problem are
\begin{align*}
W_1,W_2,Z_1,Z_2,Z_3,X_{1,1,1},X_{1,1,2},X_{1,2,1},X_{1,2,2},X_{2,1,1},X_{2,1,2},X_{2,2,1},X_{2,2,2}.
\end{align*}
\end{itemize}

\subsection{The Dependency Reduction}

The dependency (or implication) relation, e.g., the one given in (\ref{eqn:implication}), can be included in the optimization problem in different ways. The first option, which is the simplest, is to include these equality constraints directly in the set of constraints of the LP. There is however another method. Observe that since the two entropy values are equal, we can simply represent them using the same LP variable, instead of generating two different LP variables then insisting that they are of the same value. This helps reduce the number of LP variables in the problem. In our two working examples, the dependence relations are as follows:
\begin{itemize}
\item The regenerating code problem: the relations are the following 
\begin{align*}
&H(S_{1,2},S_{1,3},S_{1,4}|W_1)=0\\
&H(S_{2,1},S_{2,3},S_{2,4}|W_2)=0\\
&H(S_{3,1},S_{3,2},S_{3,4}|W_3)=0\\
&H(S_{4,1},S_{4,2},S_{4,3}|W_4)=0\\
&H(W_1|S_{2,1},S_{3,1},S_{4,1})=0\\
&H(W_2|S_{1,2},S_{3,2},S_{4,2})=0\\
&H(W_3|S_{1,3},S_{2,3},S_{4,3})=0\\
&H(W_4|S_{1,4},S_{2,4},S_{3,4})=0.
\end{align*}
This dependence structure can also be represented as a graph shown in Fig. \ref{fig:dependence}. In this graph, a given node (random variable) is a function of others random variables with an incoming edge.
\begin{figure}
\centering
\includegraphics[width=0.75\textwidth]{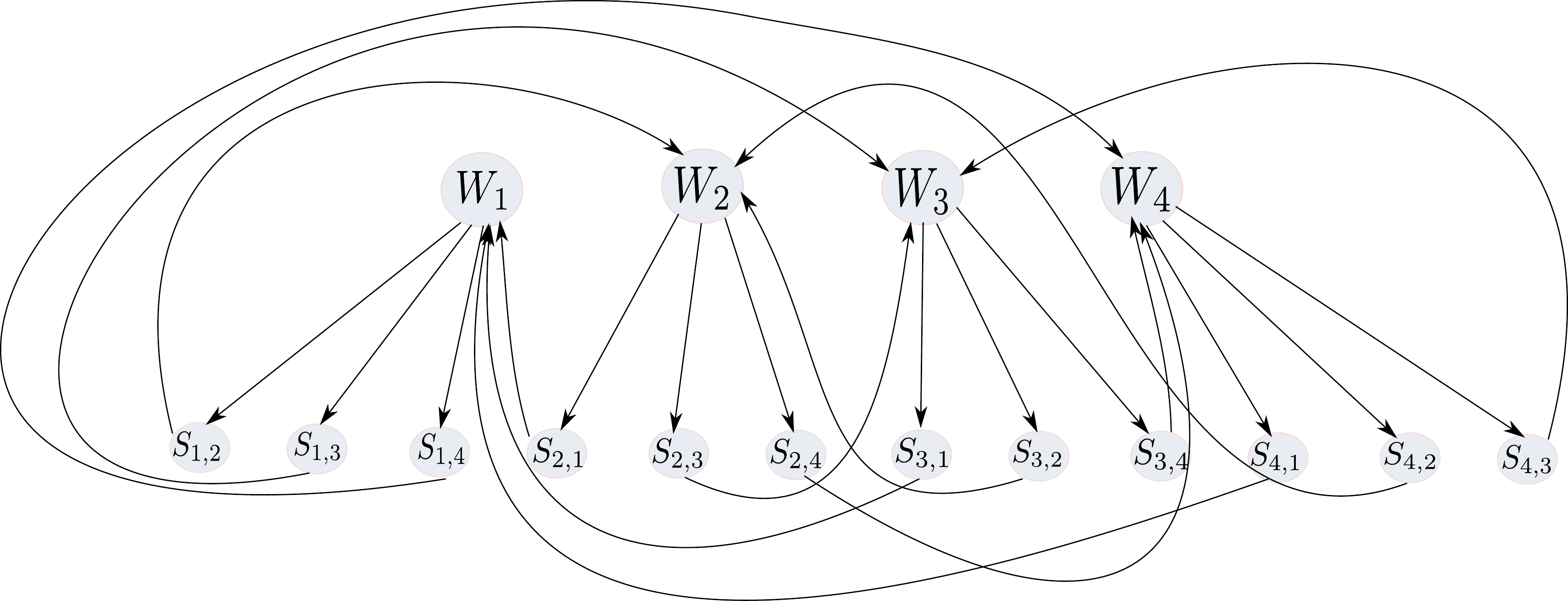}
\caption{The dependence graph for the regenerating code problem with $(n,k,d)=(4,3,3)$. \label{fig:dependence}}
\end{figure}
\item The caching problem: the relations are the following
\begin{align*}
&H(Z_1,Z_2,Z_3,X_{1,1,1},X_{1,1,2},X_{1,2,1},X_{1,2,2},X_{2,1,1},X_{2,1,2},X_{2,2,1},X_{2,2,2}|W_1,W_2)=0\\
&H(W_1|Z_1,X_{1,1,1})=0\\
&H(W_1|Z_2,X_{1,1,1})=0\\
&H(W_1|Z_3,X_{1,1,1})=0\\
&H(W_1|Z_1,X_{1,1,2})=0\\
&H(W_1|Z_2,X_{1,1,2})=0\\
&H(W_2|Z_3,X_{1,1,2})=0\\
&H(W_1|Z_1,X_{1,2,1})=0\\
&H(W_2|Z_2,X_{1,2,1})=0\\
&H(W_1|Z_3,X_{1,2,1})=0\\
&H(W_1|Z_1,X_{1,2,2})=0\\
&H(W_2|Z_2,X_{1,2,2})=0\\
&H(W_2|Z_3,X_{1,2,2})=0\\
&H(W_2|Z_1,X_{2,1,1})=0\\
&H(W_1|Z_2,X_{2,1,1})=0\\
&H(W_1|Z_3,X_{2,1,1})=0\\
&H(W_2|Z_1,X_{2,1,2})=0\\
&H(W_1|Z_2,X_{2,1,2})=0\\
&H(W_2|Z_3,X_{2,1,2})=0\\
&H(W_2|Z_1,X_{2,2,1})=0\\
&H(W_2|Z_2,X_{2,2,1})=0\\
&H(W_1|Z_3,X_{2,2,1})=0\\
&H(W_2|Z_1,X_{2,2,2})=0\\
&H(W_2|Z_2,X_{2,2,2})=0\\
&H(W_2|Z_3,X_{2,2,2})=0.
\end{align*}
\end{itemize}

\subsection{The Symmetry Reduction}

In many problems, there are certain symmetry relations present. Such symmetry relations are usually a direct consequence of the structure of the information systems. Often it is \textbf{without loss of optimality} to consider only codes with a specific symmetric structure. In our two working examples, the symmetry relations are as follows.
\begin{itemize}
\item Exchanging the coding functions for different storage nodes. For example, if we simply let node 2 store the content for node 1, and also exchange their other functions, the result is another code that can fulfill the same task as before this exchange. Mathematically, we can represent the symmetry relation using permutations of all the random variables, where each row indicates a permutation as follows:
\begin{align*}
W_1,W_2,W_3,W_4,S_{1,2},S_{1,3},S_{1,4},S_{2,1},S_{2,3},S_{2,4},S_{3,1},S_{3,2},S_{3,4},S_{4,1},S_{4,2},S_{4,3}\\
W_1,W_2,W_4,W_3,S_{1,2},S_{1,4},S_{1,3},S_{2,1},S_{2,4},S_{2,3},S_{4,1},S_{4,2},S_{4,3},S_{3,1},S_{3,2},S_{3,4}\\
W_1,W_3,W_2,W_4,S_{1,3},S_{1,2},S_{1,4},S_{3,1},S_{3,2},S_{3,4},S_{2,1},S_{2,3},S_{2,4},S_{4,1},S_{4,3},S_{4,2}\\
W_1,W_4,W_3,W_2,S_{1,4},S_{1,3},S_{1,2},S_{4,1},S_{4,3},S_{4,2},S_{3,1},S_{3,4},S_{3,2},S_{2,1},S_{2,4},S_{2,3}\\
W_1,W_3,W_4,W_2,S_{1,3},S_{1,4},S_{1,2},S_{3,1},S_{3,4},S_{3,2},S_{4,1},S_{4,3},S_{4,2},S_{2,1},S_{2,3},S_{2,4}\\
W_1,W_4,W_2,W_3,S_{1,4},S_{1,2},S_{1,3},S_{4,1},S_{4,2},S_{4,3},S_{2,1},S_{2,4},S_{2,3},S_{3,1},S_{3,4},S_{3,2}\\
W_2,W_1,W_3,W_4,S_{2,1},S_{2,3},S_{2,4},S_{1,2},S_{1,3},S_{1,4},S_{3,2},S_{3,1},S_{3,4},S_{4,2},S_{4,1},S_{4,3}\\
W_2,W_4,W_3,W_1,S_{2,4},S_{2,3},S_{2,1},S_{4,2},S_{4,3},S_{4,1},S_{3,2},S_{3,4},S_{3,1},S_{1,2},S_{1,4},S_{1,3}\\
W_2,W_1,W_4,W_3,S_{2,1},S_{2,4},S_{2,3},S_{1,2},S_{1,4},S_{1,3},S_{4,2},S_{4,1},S_{4,3},S_{3,2},S_{3,1},S_{3,4}\\
W_2,W_4,W_1,W_3,S_{2,4},S_{2,1},S_{2,3},S_{4,1},S_{4,2},S_{4,3},S_{1,2},S_{1,4},S_{1,3},S_{3,2},S_{3,4},S_{3,1}\\
W_2,W_3,W_1,W_4,S_{2,3},S_{2,1},S_{2,4},S_{3,2},S_{3,1},S_{3,4},S_{1,2},S_{1,3},S_{1,4},S_{4,2},S_{4,3},S_{4,1}\\
W_2,W_3,W_4,W_1,S_{2,3},S_{2,4},S_{2,1},S_{3,2},S_{3,4},S_{3,1},S_{4,2},S_{4,3},S_{4,1},S_{1,2},S_{1,3},S_{1,4}\\
W_3,W_2,W_1,W_4,S_{3,2},S_{3,1},S_{3,4},S_{2,3},S_{2,1},S_{2,4},S_{1,3},S_{1,2},S_{1,4},S_{4,3},S_{4,2},S_{4,1}\\
W_3,W_2,W_4,W_1,S_{3,2},S_{3,4},S_{3,1},S_{2,3},S_{2,4},S_{2,1},S_{4,3},S_{4,2},S_{4,1},S_{1,3},S_{1,2},S_{1,4}\\
W_3,W_1,W_2,W_4,S_{3,1},S_{3,2},S_{3,4},S_{1,3},S_{1,2},S_{1,4},S_{2,3},S_{2,1},S_{2,4},S_{4,3},S_{4,1},S_{4,2}\\
W_3,W_1,W_4,W_2,S_{3,1},S_{3,4},S_{3,2},S_{1,3},S_{1,4},S_{1,2},S_{4,3},S_{4,1},S_{4,2},S_{2,3},S_{2,1},S_{2,4}\\
W_3,W_4,W_1,W_2,S_{3,4},S_{3,1},S_{3,2},S_{4,3},S_{4,1},S_{4,2},S_{1,3},S_{1,4},S_{1,2},S_{2,3},S_{2,4},S_{2,1}\\
W_3,W_4,W_2,W_1,S_{3,4},S_{3,2},S_{3,1},S_{4,3},S_{4,2},S_{4,1},S_{2,3},S_{2,4},S_{2,1},S_{1,3},S_{1,4},S_{1,2}\\
W_4,W_2,W_3,W_1,S_{4,2},S_{4,3},S_{4,1},S_{2,4},S_{2,3},S_{2,1},S_{3,4},S_{3,2},S_{3,1},S_{1,4},S_{1,2},S_{1,3}\\
W_4,W_2,W_1,W_3,S_{4,2},S_{4,1},S_{4,3},S_{2,4},S_{2,1},S_{2,3},S_{1,4},S_{1,2},S_{1,3},S_{3,4},S_{3,2},S_{3,1}\\
W_4,W_1,W_3,W_2,S_{4,1},S_{4,3},S_{4,2},S_{1,4},S_{1,3},S_{1,2},S_{3,4},S_{3,1},S_{3,2},S_{2,4},S_{2,1},S_{2,3}\\
W_4,W_1,W_2,W_3,S_{4,1},S_{4,2},S_{4,3},S_{1,4},S_{1,2},S_{1,3},S_{2,4},S_{2,1},S_{2,3},S_{3,4},S_{3,1},S_{3,2}\\
W_4,W_3,W_1,W_2,S_{4,3},S_{4,1},S_{4,2},S_{3,4},S_{3,1},S_{3,2},S_{1,4},S_{1,3},S_{1,2},S_{2,4},S_{2,3},S_{2,1}\\
W_4,W_3,W_2,W_1,S_{4,3},S_{4,2},S_{4,1},S_{3,4},S_{3,2},S_{3,1},S_{2,4},S_{2,3},S_{2,1},S_{1,4},S_{1,3},S_{1,2}
\end{align*}
Note that when we permute the storage contents $(W_1,W_2,W_3,W_4)$ the corresponding repair information needs to be permuted accordingly. The 24 permutations clearly form a permutation group. With this representation, we can take any subset of the columns, and the collections of the random variables in each row in these columns will have the same entropy in the corresponding symmetric code. For example, if we take $2,9,15$ columns, then we have that
\begin{align*}
H(W_1,S_{2,1},S_{4,1})=H(W_1,S_{2,1},S_{3,1})=H(W_1,S_{3,1},S_{4,1})=...
\end{align*}
There are a total of $2^{17}-1$ subset of columns, and they each will induce a set of equality relations, but some of them may be equivalent. For a more rigorous discussion of this symmetry relation, the readers can refer to \cite{Tian:JSAC13,ZhangTian:17TCOM}.
\item Similarly, in the caching problem, there are two types of symmetry relations. The first is to exchange the coding functions for each users, and the second is to exchange the operation on different files. Intuitively, the first one is due to a permutation of the users, and the second due to the permutation of the files. As a consequence, we have the following permutations that form a group:
\begin{align*}
W_1, W_2, Z_1, Z_2, Z_3, X_{1,1,1},X_{1,1,2},X_{1,2,1},X_{1,2,2},X_{2,1,1},X_{2,1,2},X_{2,2,1},X_{2,2,2}\\
W_2, W_1, Z_1, Z_2, Z_3, X_{2,2,2},X_{2,2,1},X_{2,1,2},X_{2,1,1},X_{1,2,2},X_{1,2,1},X_{1,1,2},X_{1,1,1}\\
W_1, W_2, Z_2, Z_1, Z_3, X_{1,1,1},X_{1,1,2},X_{2,1,1},X_{2,1,2},X_{1,2,1},X_{1,2,2},X_{2,2,1},X_{2,2,2}\\
W_2, W_1, Z_2, Z_1, Z_3, X_{2,2,2},X_{2,2,1},X_{1,2,2},X_{1,2,1},X_{2,1,2},X_{2,1,1},X_{1,1,2},X_{1,1,1}\\
W_1, W_2, Z_3, Z_2, Z_1, X_{1,1,1},X_{2,1,1},X_{1,2,1},X_{2,2,1},X_{1,1,2},X_{2,1,2},X_{1,2,2},X_{2,2,2}\\
W_2, W_1, Z_3, Z_2, Z_1, X_{2,2,2},X_{1,2,2},X_{2,1,2},X_{1,1,2},X_{2,2,1},X_{1,2,1},X_{2,1,1},X_{1,1,1}\\
W_1, W_2, Z_1, Z_3, Z_2, X_{1,1,1},X_{1,2,1},X_{1,1,2},X_{1,2,2},X_{2,1,1},X_{2,2,1},X_{2,1,2},X_{2,2,2}\\
W_2, W_1, Z_1, Z_3, Z_2, X_{2,2,2},X_{2,1,2},X_{2,2,1},X_{2,1,1},X_{1,2,2},X_{1,1,2},X_{1,2,1},X_{1,1,1}\\
W_1, W_2, Z_2, Z_3, Z_1, X_{1,1,1},X_{2,1,1},X_{1,1,2},X_{2,1,2},X_{1,2,1},X_{2,2,1},X_{1,2,2},X_{2,2,2}\\
W_2, W_1, Z_2, Z_3, Z_1, X_{2,2,2},X_{1,2,2},X_{2,2,1},X_{1,2,1},X_{2,1,2},X_{1,1,2},X_{2,1,1},X_{1,1,1}\\
W_1, W_2, Z_3, Z_1, Z_2, X_{1,1,1},X_{1,2,1},X_{2,1,1},X_{2,2,1},X_{1,1,2},X_{1,2,2},X_{2,1,2},X_{2,2,2}\\
W_2, W_1, Z_3, Z_1, Z_2, X_{2,2,2},X_{2,1,2},X_{1,2,2},X_{1,1,2},X_{2,2,1},X_{2,1,1},X_{1,2,1},X_{1,1,1}
\end{align*}
For a more detailed discussion on this symmetry relation, the readers can refer to \cite{Tian:16Computer,ZhangTian:17TCOM}.
\end{itemize}

Three cautions should be given at this point when working with symmetry relation:
\begin{itemize}
\item The symmetry reduction is valid in the entropy LP, when it is without loss of optimality to consider only the corresponding symmetry codes. Such assertion is usually not too difficult to make with a concrete problem in mind, but it can be less clear when the problem is abstract.
\item The precise permutations can be subtle. For example, observe in the caching example, the difference between the first row and the eleven-th row is induced by the permutation of the user contents into $(Z_3,Z_1,Z_2)$, and as a result, the indices of the $X_{i,j,k}$ variables will change. However, the precise way it is changed to is less intuitive and needs to be carefully worked out. In fact, when there are more than two messages, the permutation is even more complicated; readers are referred to \cite{Tian:16Computer} for a thorough discussion. 
\item For the purpose of deriving outer bounds, it is valid to ignore the symmetry relation altogether, or consider only part of the symmetry relation, as long as the remaining permutations still form a group. For example, in the caching problem if we only consider the symmetry induced by exchange the two messages, then we only have the first 2  rows instead of the full 12 rows of permutations. Omitting some permutations means less reduction in the LP scale, but does not invalidate the computed bounds.
\item Admitted, representing the symmetry relation using the above permutation representation is not the most concise approach, and there exist mathematically precise and concise language to specify such structure. We choose this approach in the toolbox because of its simplicity and universality. Moreover, advanced representations can be used to produce this representation easily. 
\end{itemize}

\section{Four Streamlined Functionalities}

There are four main functionalities that we implemented in this toolbox and are ready for use. In this section, we provide an overview of the usages. 

\subsection{Bounding Plane Optimization and Queries}

\begin{figure}
\centering
\includegraphics[width=0.5\textwidth]{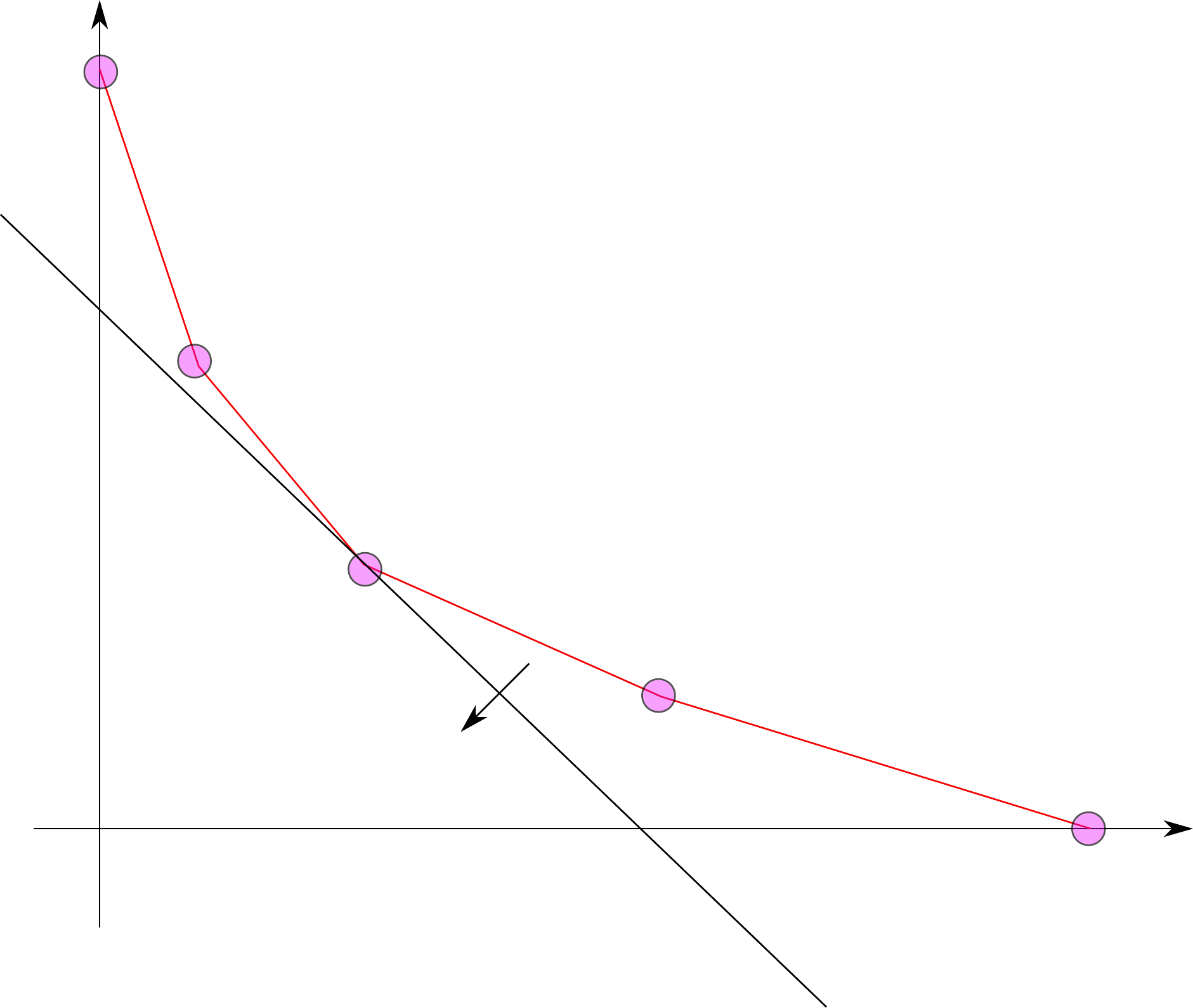}
\caption{A fixed direction bounding plane and the tradeoff region computation.\label{fig:region}}
\end{figure}

In this case, the objective function is fixed, and the optimal solution gives an outer bound of a specific linear combination of several information measures or relevant quantities. Fig. \ref{fig:region} illustrates the idea for this specific functionality, where we wish to find a lower bound of the given direction for the given optimal tradeoff shown in red.

 Let us again consider the two working examples.
\begin{itemize}
\item If the simple sum  of the storage cost $\alpha$ and repair cost $\beta$, i.e., $\alpha+\beta$, needs to be lower bounded in the regenerating code problem, we can let the objective function be given as
\begin{align*}
H(W_1)+H(S_{1,2}) 
\end{align*}
and then minimize it. The optimal value will be a lower bound, which in this case turns out to be $5/8$. Note that by taking advantage of the symmetry, the objective function set up above indeed specifies sum of the storage cost and repair cost for any storage and repair transmission. 
\item If we wish to lower bound the simple sum of memory and rate in the coded caching problem, the situation is somewhat sutble. Note that the rate $R$ is a lower bound on the entropy $H(X_{1,1,1})$ and $H(1,2,2)$, however, the symmetry relation does not imply that  $H(X_{1,1,1})=H(X_{1,2,2})$. For this case, we can introduce an additional LP variable $R$, and add the constraints that
\begin{align*}
H(X_{1,1,1})\leq R\\
H(X_{1,2,2})\leq R,
\end{align*}
and then set the objective function to be 
\begin{align*}
H(Z_1)+R,
\end{align*}
from which the minimum value is an lower bound on the simple sum of memory and rate in this setting.
\end{itemize}

In addition to simply computing the supporting hyperplane, the toolbox allows the user to extract useful information on the optimal solution. Particularly, the user can probe for the values of certain information measures. For example, in the case above for coded caching, we may be interested in the value of $I(Z_1;W_1)$, which essentially reveals the amount of information regarding $W_1$ that is stored in $Z_1$ in an uncoded form.

\subsection{Tradeoff and Convex Hull Computation}

In many cases, instead of bounding a fix rate combination, we are interested in the tradeoff of several quantities, in fact, most time the optimal tradeoff between two quantities; see Fig. \ref{fig:region} again for an illustration. The two working examples both belong to this case. 

Since the constrained set in the LP is a polytope, the resulting outer bound to the optimal tradeoff will be a piece-wise linear bound. A naive strategy is to trace the boundary by sampling points on a sufficiently dense grid. However, this approach is  time consuming and not accurate. Instead the calculation of this piece-wise linear outer bound is equivalent to computing the projection of a convex polytope, for which Lassez's algorithm in fact provides a method to complete the task efficiently. We implemented Lassez's algorithm for the projection on to two-dimensional space in this toolbox. A more detailed description of this algorithm can be found in \cite{Lassez:92}, and the specialization used in the program can be found in \cite{Tian:16Computer}. 

\subsection{Duality and Computer-Generated Proof}

After identifying a valid outer bound, we sometimes wish to find a proof for this bound. In fact, even if the bound is optimal, or it is a only hypothesis bound, we may still attempt to prove it. For example, in the regenerating code problem, we have
\begin{align}
H(W_1)+H(S_{1,2})\geq \frac{5}{8}. 
\end{align}
How can we prove this inequalities? It is clear from the LP duality that this inequality is a weighted sum of the individual constraints in the LP. Thus as long as we find one such weighted sum, we can then write down a chain of inequalities directly by combining these inequalities one by one; for a more detailed discussion, see \cite{Tian:JSAC13,Tian:16Computer,ho2014proving,li2016multilevel}.  In this toolbox, this task of finding a proof is implemented, which will print out a computed weight sum. Note, however, such a proof is not unique, and the proof such found may lack a structure desirable for further generalization, and subsequent processing may be necessary for this purpose.

\subsection{Sensitivity Analysis}

At the computed optimal value, we can probe for the range of certain information measures such that forcing them to be in these values does not change the value of the optimal solution. Consider the quantity $I(Z_1;W_1)$ in the caching problem, it may be possible for it take values between $[0.2,0.4]$ without changing the optimal value of the original optimization problem. On the other hand, if it can only take value $0.2$, then this suggests if a code to achieve this optimal value indeed exists, it must have this amount of uncoded information regarding $W_1$ stored in $Z_1$. This information can be valuable in reverse-engineering optimal codes; see \cite{Tian:16Computer} for discussion of such usage. 

\section{Software Dependence and Computation Pipeline}

The toolbox needs a linear programming solver as the backend. There are two options currently allowed in the toolbox: Cplex \cite{Cplex} or Gurobi \cite{Gurobi}. Both solvers are known to perform extremely well for large scale linear programs, and offer free license for academic use (but not commercial use). There is no need to have both solvers, and installing one package will be sufficient. A make file is given in the package, and a visual C++ solution configuration (VC2019) is also given. The only place that we need to pay attention is to link the solver library correctly. The toolbox otherwise has no dependence on third party packages that must be installed. The instruction to install the toolbox can be found inside the online source repository. 

\begin{figure}[t]
\centering
\includegraphics[width=0.9\textwidth]{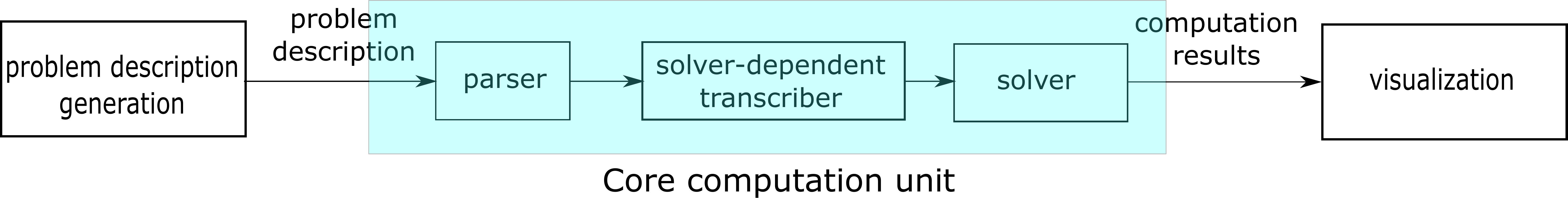}
\caption{The computation pipeline. \label{fig:pipeline}}
\end{figure}

The toolbox has a relatively sophisticated front-end parser, and this may be one of the most notable features different from existing efforts \cite{ITIP,XITIP}. The parser will parse a problem description file into an inherent data structure, after certain error and syntax checking. Before the solver can be called, the problem must also be transcribed into the linear program using the reduction technique we previous discussed. The parser is not solver dependent, but the transcriber is solver dependent, as the transcribing process can be done more efficiently by utilizing the solver APIs. After the problem is transcribed to an LP suitable for the chosen solver, the solver can then be called to solve the program. Depending on the mode of computation, the results are then translated and print to output. The core computation unit is illustrated in Fig. \ref{fig:pipeline} in the shaded blue area.

The core computation can be embedded into a more sophisticated computer-aided exploration system. In order to illustrate such usage, in this toolbox we include a python script that can complete the process of PD generation$\rightarrow$computation$\rightarrow$visualization process for several well-known problems of current interest (e.g., coded caching and private information retrieval). Given the ease of developing functions using python, it is our hope that more advanced functions can be developed in the near future.

\section{Syntax for Problem Description Files}

The program will read a problem description file (a plain text file), and the desired computed bounds or proof will be produced without further user intervention. In this section, we provide the details of the syntax of the problem description file.
The input problem description files must include the characters \code{PD} (which stand for ``problem description"), followed by a JSON detailing the problem description. The JSON can either be on the same line as \code{PD} or start on the next line, but it cannot begin on the same line as \code{PD} and then continue on the next line.

\subsection{Keys in \code{PD} JSON}
There are a total of 12 keys allowed in the \code{PD} JSON in the description file: 
\begin{center}
\code{RV}, \code{AL}, \code{O}, \code{D}, \code{I}, \code{S}, \code{BC}, \code{BP}, \code{QU}, \code{SE}, \code{CMD}, and \code{OPT}.
\end{center}
The program will assume the problem to be minimization, i.e., to find a lower bound for certain information quantity. If instead we need to derive an upper bound, then the objective function will need to be written in its negated form.

    \subsubsection{\code{RV}: Random Variables}

    The \code{RV} key is used to add random variables, which must be alphanumeric strings not beginning with a number. \code{RV}'s value must be a JSON array of strings, even if there's only one random variable. For example, the following are allowed:
    \begin{itemize}
        \item \code{"RV" : ["X","YQ123","Z","var3"]}
        \item \code{"RV" : ["Z"]}
    \end{itemize}
    The following are not allowed:
    \begin{itemize}
        \item \code{"RV" : "X"~~~~~~// Not an array}
        \item \code{"RV" : X~~~~~~~~// Not an array, no quotes}
        \item \code{"RV" : [X]~~~~~~// No quotes}
        \item \code{"RV" : [X,Y]~~~~// No quotes}
    \end{itemize}

    \subsubsection{\code{AL}: Additional LP Variables}

    The \code{AL} key is used to add additional linear programming variables, which must be alphanumeric strings not beginning with a number. \code{AL}'s value must be a JSON array of strings, even if there's only one variable. For example, the following are allowed:
    \begin{itemize}
        \item \code{"AL" : ["A","B","AB123"]}
        \item \code{"AL" : ["C"]}
    \end{itemize}
    
    \subsubsection{\code{O}: Objective Function}

    The \code{O} key is used to add the objective function. Its value should be a string written as expected. Spacing is ignored, though recommended for human readability. For example, the following is allowed:

    \begin{itemize}
        \item \code{"O" : "A + 5B + H(X) - 2H(X,YQ123|Z)"}
    \end{itemize}

    \subsubsection{\code{D}: Dependencies}

    The \code{D} key is used to add dependency relations. Its value must be an array of JSON objects, each of which specifies a dependency. Each dependency JSON object should have a \code{dependent} key and a \code{given} key, each of which has a value which is an array of random variables. For example, the dependencies 
    \begin{itemize}
        \item \code{H(X,YQ123|Z) = 0}
    \end{itemize}
    would be represented by
    \begin{itemize}
        \item \code{"D" : [ \newline
                    \tab\tab \{"dependent" : ["X","YQ123"] , "given" : ["Z"]\} , \newline
                    \tab ]}
    \end{itemize}

    \subsubsection{\code{I}: Independence}

    The \code{I} key  is used to add independence and conditional independence relations. Its value must be an array of JSON objects, each of which specifies an independence. Each independence JSON object should have an \code{independent} key and a \code{given} key, each of which has a value which is an array of random variables. For example, the independences
    \begin{itemize}
        \item \code{I(X;YQ123|Z) = 0}
        \item \code{I(X;Z) = 0}
    \end{itemize}
    would be represented by
    \begin{itemize}
        \item \code{"I" : [ \newline
                    \tab\tab \{"independent" : ["X","YQ123"] , "given" : ["Z"]\} , \newline
                    \tab\tab \{"independent" : ["X","Z"] , "given" : []\} \newline
                    \tab ]}
    \end{itemize}
    
    Independence relation can be directly represented using the corresponding information theoretic equalities, and thus can be included in the problem description using the \code{BC} key. Even the case that three random variables $X,Y,W$ are independent given a fourth one $Z$, it can be represented as 
    \begin{align*}
    H(X|Z)+H(Y|Z)+H(W|Z)-H(X,Y,W|Z)=0
    \end{align*}    
This is the recommended usage. The key \code{I} is retained in this version for legacy reasons, but may be removed in future versions. 

    \subsubsection{\code{S}: Symmetries}

    The \code{S} key is used to add symmetries. Its value must be an array of arrays of random variables, with each inner array containing every random variable exactly once. For example, assuming the set of all random variables is \code{["X1","X2","X3","X4"]}, The following is a valid set of symmetries:
    \begin{itemize}
        \item \code{"S" : [ \newline
                    \tab[1.5cm] ["X1","X2","X3","X4"], \newline
                    \tab[1.5cm] ["X1","X3","X2","X4"], \newline
                    \tab[1.5cm] ["X1","X2","X4","X3"], \newline
                    \tab[1.5cm] ["X1","X3","X4","X2"], \newline
                    \tab[1.5cm] ["X1","X4","X2","X3"], \newline
                    \tab[1.5cm] ["X1","X4","X3","X2"] \newline
                    \tab ]}
    \end{itemize}

    \subsubsection{\code{BC}: Constant Bounds}

    The \code{BC} key is used to add constant bounds. Its value should be an array of strings of (in)equalities. The right-hand side must be a number. For example:
    \begin{itemize}
        \item \code{"BC" : [ \newline
                    \tab \tab "H(X,Z) + 2A $<=$ 4" , \newline
                    \tab \tab "H(YQ123) $<=$ 12" \newline
                    \tab ]}
    \end{itemize}

    \subsubsection{\code{BP}: Bounds to Prove}
    The \code{BP} key is used to add bounds to prove. Its value should be an array of strings of (in)equalities. The right-hand side must be a number. For example:
    \begin{itemize}
        \item \code{"BP" : [ \newline
                    \tab \tab "H(X,Z) + 2A $<=$ 4" , \newline
                    \tab \tab "H(YQ123) $<=$ 12" \newline
                    \tab ]}
    \end{itemize}
    
    \subsubsection{\code{QU}: Queries on the Optimal Solution}
    
    The key \code{QU} allows the specification of quantities to retrieve the corresponding values at the optimal solution. This implies that it is only meaningful when the bounding plane computation is performed, but not for sensitivity analysis, the convex hull, or the prove computation. For example, the following form is allowed
    \begin{itemize}
    \item \code{"QU": ["2I(X;Z|YQ123)", "H(X|YQ123)"]}
    \end{itemize}
    
    \subsubsection{\code{SE}: Sensitivity Analysis}
    
    The key \code{SE} allows the specification of quantities to perform sensitivity analysis.  Its value should be an array of strings of linear combination of information measures and the additional LP variables. For example, the following form is allowed
    \begin{itemize}
    \item \code{"SE": ["A", "B", "2I(Z;X|YQ123)+H(YQ123|X)+A"]}
    \end{itemize}
    Note here \code{A} and \code{B} are two additional LP variables, but not random variables in the PD. 

    \subsubsection{\code{CMD} and \code{OPT}: Commands and Options}
    The keys \code{CMD} and \code{OPT} are identical and are both included for legacy reasons. Their value should be an array of commands, and the values of this array can consist of \code{SER}, \code{PDC}, \code{LP\_DISP}, \code{CS}, and  \code{?}. These option will tell the parser to run the command (or compute with the given option) after parsing the rest of the \code{PD}. The meaning of these four commands or options are as follows:
    \begin{itemize}
    \item \code{SER [-a|-t file]}: The program will print out the problem description state, followed by an optional file to output to (with an append or truncate flag). If no output file specified, this will output to standard out. 
    \item \code{PDC}: The program will print the current state in the classic (version 0.1) problem description format. This is an effective approach to check whether the PD file has all the necessary specifications as intended.
    \item \code{CS}: The program will perform a verification whether the symmetries entered are valid (is a permutation group). This is an effective approach to confirm the symmetry relation generated for the specific problem is as intended.
    \item \code{LP\_DISP}: This tells the LP solver to display the usual LP intermediate computation results; if this is not set, the display of the LP solver will be suppressed. This feature can be useful to avoid excessive information being printed to the terminal window. In particularly, for sensitivity analysis on multiple quantities, we recommend suppressing the LP solver output.
    \item \code{?}: The usage format of the command and PD file format will be printed. Note that this command will also be called in default if the command line arguments are invalid.
    \end{itemize}
   For example, the following options can be included in the PD JSON object.    
   \begin{itemize}
   \item \code{"OPT": ["PDC","CS","LP\_DISP"]} 
   \end{itemize}

    \subsection{Commands outside of \code{PD} JSON}

    There are a few commands that can be placed in the problem description file, and comments are allowed. To comment, simply start the line with a \code{\#}. (Note that comments are not allowed within a JSON).  The commands \code{SER}, \code{PDC}, \code{CS}, and \code{?} can also be invoked directly outside of \code{PD} JSON. This means that the options given above can also be included as follows outside the PD object in the PD file:
    \begin{itemize}
   \item \code{CMD ["PDC","CS","LP\_DISP"]}
   \item \code{PDC}
   \item \code{CS}
   \end{itemize}
   However note that in the direct command mode (e.g., the latter two cases above), they should be given after the \code{PD} JSON, since otherwise the \code{PD} would not been been read, and the commands are essentially called on an empty \code{PD}.
    
    The following features are experimental, and are not recommended to standard users as they are less extensively tested. The command \code{DESER} allows the program to enter PD from a sequence of files or even from a terminal in an interactive mode:    
    \begin{itemize}
        \item \code{DESER file1 [file2 ...]}: To read in commands from a second file, use the command \code{DESER} following by at least one file name. 
        \item \code{Q}: This will quit the interactive input mode.
        \item \code{DESTROY}: In the interactive mode, this will clean all existing result, and start a new PD completely. 
    \end{itemize}
 
   In fact, even for the problem description portion, we may use the PD keys as commands to insert them. For example: 
\begin{itemize}
   \item {\scriptsize \code{RV ["W1","W2","W3","W4","S12","S13","S14","S21","S23","S24","S31","S32","S34","S41","S42","S43"] }} 
\end{itemize}

    \subsection{An Example Problem Description File} \label{sec:examplepd}
    Below is a  tested PD file for the regenerating code problem we previous discussed. 
    
{
\tt 
\small
\begin{verbatim}
# including the options as below is also valid 
# CMD ["PDC","CS","LP\_DISP"]
# PDC
PD
{
  "OPT": ["PDC","CS"],
  "RV": ["S12","S13","S14","S21","S23","S24","S31","S32","S34","S41","S42","S43"],
  "AL": ["A","B"],
  "O": "A+B",
  "D": [
    {"dependent":["S12","S13","S14"],"given":["S21","S31","S41"]},
    {"dependent":["S21","S23","S24"],"given":["S12","S32","S42"]},
    {"dependent":["S31","S32","S34"],"given":["S13","S23","S43"]},
    {"dependent":["S41","S42","S43"],"given":["S14","S24","S34"]} ],
  "I": [],
  "BC": [
    "H(S12,S13,S14) - A <= 0",
    "H(S12) - B <= 0",
    "H(S12,S13,S14,S21,S23,S24,S31,S32,S34,S41,S42,S43) >= 1" ],
  "BP": ["4A + 6B >= 3"],
  "SE": ["A", "B", "2I(S12;S21|S32)+H(S21|S31)+A"],
  "QU": ["A", "B", "2H(S12|S13)","-2I(S12;S21|S32)"],
  "S": [
    ["S12","S13","S14","S21","S23","S24","S31","S32","S34","S41","S42","S43"],
    ["S12","S14","S13","S21","S24","S23","S41","S42","S43","S31","S32","S34"],
    ["S13","S12","S14","S31","S32","S34","S21","S23","S24","S41","S43","S42"],
    ["S14","S13","S12","S41","S43","S42","S31","S34","S32","S21","S24","S23"],
    ["S13","S14","S12","S31","S34","S32","S41","S43","S42","S21","S23","S24"],
    ["S14","S12","S13","S41","S42","S43","S21","S24","S23","S31","S34","S32"],
    ["S21","S23","S24","S12","S13","S14","S32","S31","S34","S42","S41","S43"],
    ["S24","S23","S21","S42","S43","S41","S32","S34","S31","S12","S14","S13"],
    ["S21","S24","S23","S12","S14","S13","S42","S41","S43","S32","S31","S34"],
    ["S24","S21","S23","S42","S41","S43","S12","S14","S13","S32","S34","S31"],
    ["S23","S21","S24","S32","S31","S34","S12","S13","S14","S42","S43","S41"],
    ["S23","S24","S21","S32","S34","S31","S42","S43","S41","S12","S13","S14"],
    ["S32","S31","S34","S23","S21","S24","S13","S12","S14","S43","S42","S41"],
    ["S32","S34","S31","S23","S24","S21","S43","S42","S41","S13","S12","S14"],
    ["S31","S32","S34","S13","S12","S14","S23","S21","S24","S43","S41","S42"],
    ["S31","S34","S32","S13","S14","S12","S43","S41","S42","S23","S21","S24"],
    ["S34","S31","S32","S43","S41","S42","S13","S14","S12","S23","S24","S21"],
    ["S34","S32","S31","S43","S42","S41","S23","S24","S21","S13","S14","S12"],
    ["S42","S43","S41","S24","S23","S21","S34","S32","S31","S14","S12","S13"],
    ["S42","S41","S43","S24","S21","S23","S14","S12","S13","S34","S32","S31"],
    ["S41","S43","S42","S14","S13","S12","S34","S31","S32","S24","S21","S23"],
    ["S41","S42","S43","S14","S12","S13","S24","S21","S23","S34","S31","S32"],
    ["S43","S41","S42","S34","S31","S32","S14","S13","S12","S24","S23","S21"],
    ["S43","S42","S41","S34","S32","S31","S24","S23","S21","S14","S13","S12"] ]
}
\end{verbatim}
}

    \subsection{Basic Computation Modes: Command Line Argument} \label{sec:callingcplexcompute}

   The executable allows a few computation modes when called as follows: \\
   \\
   \code{CplexCompute pdfile [hull|random|prove|sensitivity]}   \\
   \\
  The first argument is the input PD file, and the second is the optional computation mode. There are a total of five possible modes. 
   \begin{itemize}
   \item \code{regular}: This is the default computation mode (when the option is omitted), and it computes the bounding plane value;
   \item \code{hull}: This computes the tradeoff region convex hull. The two quantities to compute the tradeoff between are extracted from the objective functions, which implies that the objective function should be one of the three cases: 1) the linear combination of two joint entropies, 2) the linear combination of one joint entropy and one additional LP variable, or  3) the linear combination of two additional LP variables. It is highly recommended to introduce additional LP variables to represent both quantities.
   \item \code{random}: This is an experimental function which using only a random set of the Shannon-type inequalties to perform the computation. It is not recommended to use at this point, except for advanced users.
   \item \code{prove}: This will prove the target information inequalities.
   \item \code{sensitivity}: Sensitivity analysis on the target information quantities will be performed. 
   \end{itemize}
   
   \subsection{Command Line Modifier}
   
   Though this not the recommended usage, the executable also allows flexible modifiers through the command line: \\
   \\
   \code{CplexCompute pdfile [hull|random|prove|sensitivity] [cmd-line-modifier-1 ...]}   \\
   \\
The input file can be modified by using optional command-line modifiers. Each command-line modifier will need to be a JSON whose keys are a subset of the following: \code{RV}, \code{AL}, \code{O}, \code{D}, \code{I}, \code{S}, \code{BC}, \code{BP}, \code{+RV}, \code{+AL}, \code{+D}, \code{+I}, \code{+S}, \code{+BC}, \code{+BP}, \code{+CMD}, \code{+OPT}. (Note the exclusion of \code{+O}, \code{CMD}, and \code{OPT}.) These are the same keys allowed in the PD JSON (with the exclusion of \code{CMD} and \code{OPT}), and those keys preceded by a \code{+} (with the exclusion of \code{+O}). The value for each of these keys must be the same as for the PD JSON equivalent: a JSON array containing elements in the same format as in the PD JSON (with the exception that the \code{O} key can have a value which is just a string; the array isn't necessary). Keys that don't include a \code{+} sign (i.e. keys that are identical to the PD JSON keys) will replace that line from the input file, while keys that do inclue a \code{+} sign will append the item(s) from its value to the appropriate part of the problem description. The only elements of the JSON array which is the value of \code{+CMD} or \code{+OPT} are \code{SER}, \code{PDC}, and \code{CS}. Note that this is an advanced option, and it is not recommended for entry level users. 

    Here is an example with a description of what happens:\\
\\
{\small
    \code{CplexCompute inputfile.pd '{"+BC" : ["H(X) = 0", "H(Y) = 0"] ,}\\
    \code { "D" : [{"dependent" : ["A","B"] , "given" : ["C"]}]}' '{"+CMD" : ["PDC"]}'}}\\
    \begin{enumerate}
        \item \code{CplexCompute} reads \code{inputfile.pd}.
        \item \code{"H(X) = 0"} and \code{"H(Y) = 0"} are added to the \code{BC} currently stored.
        \item All dependencies in memory from \code{inputfile.pd} are deleted.
        \item The dependency in the command-line modifier becomes the only dependency.
        \item The program records that it needs to run \code{PDC} before sending data to Cplex.
    \end{enumerate}

   Note that the Windows Command Prompt handles quotes differently. For Windows Command Prompt users, each command-line modifier needs to be surrounded by double-quotes, and the inner quotes should be escaped with a backslash. The above example would be written as follows:\\
\\
{\small
    \code{CplexCompute inputfile.pd "\{\textbackslash"+BC\textbackslash" : [\textbackslash"H(X) = 0\textbackslash", \textbackslash"H(Y) = 0\textbackslash"],} \\
    \code{\textbackslash"D\textbackslash" : [\{\textbackslash"dependent\textbackslash" : [\textbackslash"A\textbackslash",\textbackslash"B\textbackslash"] , \textbackslash"given\textbackslash" : [\textbackslash"C\textbackslash"]\}]\}" "\{\textbackslash"+CMD\textbackslash" : [\textbackslash"PDC\textbackslash"]\}"}
}

\subsection{Example Command Lines and Computation Results}

Consider the command line
\begin{verbatim}
       CplexCompute PDRG4x3x3.pd
\end{verbatim}
where the problem description file is for our working example of the regenerating code problem, with the \code{LP\_DISP} and \code{CS} turned on. The output is:

\begin{center} 
{\tt \small
\begin{verbatim}
Symmetries have been successfully checked.
Total number of elements before reduction: 65536
Total number of elements after reduction: 179
Total number of constraints given to Cplex: 40862
CPXPARAM_Read_DataCheck                          1
Tried aggregator 1 time.
DUAL formed by presolve
LP Presolve eliminated 35779 rows and 3 columns.
Reduced LP has 177 rows, 5084 columns, and 17831 nonzeros.
Presolve time = 0.08 sec. (23.15 ticks)
Parallel mode: using up to 4 threads for barrier.
Number of nonzeros in lower triangle of A*A' = 5960
Using Approximate Minimum Degree ordering
Total time for automatic ordering = 0.00 sec. (0.44 ticks)
Summary statistics for Cholesky factor:
  Threads                   = 4
  Rows in Factor            = 177
  Integer space required    = 923
  Total non-zeros in factor = 12808
  Total FP ops to factor    = 1232248
 Itn      Primal Obj        Dual Obj  Prim Inf Upper Inf  Dual Inf Inf Ratio
   0   1.0000000e+01   0.0000000e+00  3.77e+04  0.00e+00  5.26e+03  1.00e+00
   1   6.1373374e+00   1.5078468e+00  2.60e+04  0.00e+00  3.34e+03  2.44e+00
   2   2.2445860e+00   1.4338756e+00  1.07e+04  0.00e+00  1.06e+03  8.08e+04
   3   1.5228039e+00   9.6993458e-01  4.73e+03  0.00e+00  1.81e+02  9.34e+01
   4   1.0830951e+00   9.9552024e-01  7.79e+02  0.00e+00  1.94e+01  1.01e+03
   5   1.0262480e+00   9.8688399e-01  3.25e+02  0.00e+00  2.65e+00  7.06e+03
   6   1.0237260e+00   9.7472519e-01  3.14e+02  0.00e+00  1.78e+00  6.55e+03
   7   9.5978567e-01   9.5337070e-01  4.00e+01  0.00e+00  2.14e-01  5.25e+04
   8   9.0444925e-01   8.9317974e-01  2.54e+01  0.00e+00  1.26e-01  6.17e+04
   9   8.0010814e-01   8.3512233e-01  7.88e+00  0.00e+00  7.35e-02  8.41e+04
  10   7.6631321e-01   7.7048934e-01  5.74e+00  0.00e+00  4.38e-02  1.33e+05
  11   7.0912848e-01   7.0989975e-01  2.21e+00  0.00e+00  1.64e-02  3.38e+05
  12   6.5432023e-01   6.5701740e-01  6.27e-01  0.00e+00  5.20e-03  9.72e+05
  13   6.2554618e-01   6.2728184e-01  1.45e-02  0.00e+00  3.91e-04  1.33e+07
  14   6.2499995e-01   6.2500055e-01  1.44e-06  0.00e+00  1.13e-07  5.06e+10
  15   6.2500000e-01   6.2500000e-01  3.40e-10  0.00e+00  1.77e-11  4.65e+14
Barrier time = 0.16 sec. (36.63 ticks)

Total time on 4 threads = 0.16 sec. (36.63 ticks)

******************************************************************
Optimal value for A + B = 0.625000.
Queried values:
A                         = 0.37500
B                         = 0.25000
2H(S12|S13)               = 0.25000
-2I(S12;S21|S32)          = -0.25000
******************************************************************
\end{verbatim}
}
\end{center}

The first part of the output is various information about the problem and the corresponding check and verification result. If any verification fails, the program will exit. 
The next part of the output content is the computation progress given by the chosen solver. The last star-separated segment means that we have a bound
\begin{align*}
\alpha+\beta\geq 0.625,
\end{align*}
The queried quantities are also shown in the part and it can be seen $(\alpha,\beta)$ in the LP optimal solution are $(0.375,0.25)$, together with the values of two other information measures. 

If instead we run the following command line
\begin{verbatim}
       CplexCompute PDRG4x3x3.pd hull
\end{verbatim}
with the \code{LP\_DISP} option removed, then the result shows: 

\begin{center} 
{\tt \small
\begin{verbatim}
Total number of elements before reduction: 65536
Total number of elements after reduction: 179
Total number of constraints given to Cplex: 40862
New point (0.333333, 0.333333).
New point (0.500000, 0.166667).
New point (0.375000, 0.250000).


List of found points on the hull:
(0.333333, 0.333333).
(0.375000, 0.250000).
(0.500000, 0.166667).
End of list of found points.
\end{verbatim}
}
\end{center}
In this case, the solver output is suppressed, and it is seen that the three $(\alpha,\beta)$ pairs are the corner points of the lower convex hull of the tradeoff. Occasionally, certain non-corner points on the boundary will also be included in the list, but this does not invalid the hull. In other words, every corner points of the convex hull will be in this list, but not all the points in this list will be a corner point. The non-corner points can be addressed easily in subsequent processing.  

If we run the following command: 
\begin{verbatim}
       CplexCompute PDRG4x3x3.pd prove
\end{verbatim}
we will have the following result:
\begin{center} 
{\tt \scriptsize
\begin{verbatim}
Total number of elements before reduction: 65536
Total number of elements after reduction: 179
Total number of constraints given to Cplex: 40862
******************************************************************
LP dual value 7.000000
Proved 1-th inequality: 2A + B >= 1.
001-th inequality: weight = 2.000000     -H(W1,S24)     H(S13)     H(W2)>=0
002-th inequality: weight = 1.000000     -H(W1,W2,W3,W4)    2.0H(W1,S24)    -H(S13)>=0
003-th inequality: weight = 2.000000     -H(W2)       A>=0
004-th inequality: weight = 1.000000     -H(S13)       B>=0
005-th inequality: weight = 1.000000      H(W1,W2,W3,W4)    -1.0*>=0
******************************************************************
******************************************************************
MIP dual value 7.000000
Proved 1-th inequality using integer values: 2A + B >= 1.
001-th inequality: weight = 2.000000     -H(W1,S24)     H(S13)     H(W2)>=0
002-th inequality: weight = 1.000000     -H(W1,W2,W3,W4)    2.0H(W1,S24)    -H(S13)>=0
003-th inequality: weight = 2.000000     -H(W2)       A>=0
004-th inequality: weight = 1.000000     -H(S13)       B>=0
005-th inequality: weight = 1.000000      H(W1,W2,W3,W4)    -1.0*>=0
******************************************************************
******************************************************************
LP dual value 29.000000
Proved 2-th inequality: 4A + 6B >= 3.
001-th inequality: weight = 1.000000      H(W1,W3,S21,S41)    -H(W1,W3,S21,S23,S41)     H(W1,S24,S31,S41)    -H(W1,S24,S41)>=0
002-th inequality: weight = 3.000000     -H(W1,W3,S21,S41)    2.0H(W1,S24)    -H(W2)>=0
003-th inequality: weight = 7.000000     -H(W1,S24)     H(S13)     H(W2)>=0
004-th inequality: weight = 1.000000     -H(S13)     H(W1,S31)     H(S13,S24)    -H(W1,S23,S41)>=0
005-th inequality: weight = 1.000000     -H(W1,W2,W3,W4)    -H(W1,S31)     H(W1,W4,S21)     H(W1,S24,S41)>=0
006-th inequality: weight = 1.000000      H(W1,W3,S21,S41)    -H(W1,W4,S21)>=0
007-th inequality: weight = 1.000000     -H(W1,W2,W3,W4)     H(W1,W3,S21,S41)     H(W1,W3,S21,S23,S41)    -H(W1,S24,S31,S41)>=0
008-th inequality: weight = 1.000000     -H(W1,W2,W3,W4)     H(W1,S24)    -H(S13,S24)     H(W1,S23,S41)>=0
009-th inequality: weight = 4.000000     -H(W2)       A>=0
010-th inequality: weight = 6.000000     -H(S13)       B>=0
011-th inequality: weight = 3.000000      H(W1,W2,W3,W4)    -1.0*>=0
******************************************************************
******************************************************************
MIP dual value 29.000000
Proved 2-th inequality using integer values: 4A + 6B >= 3.
001-th inequality: weight = 1.000000      H(W1,W3,S21,S41)    -H(W1,W3,S21,S23,S41)     H(W1,S24,S31,S41)    -H(W1,S24,S41)>=0
002-th inequality: weight = 3.000000     -H(W1,W3,S21,S41)    2.0H(W1,S24)    -H(W2)>=0
003-th inequality: weight = 1.000000      H(W1,W3,S21,S41)    -H(W1,W4,S21)>=0
004-th inequality: weight = 7.000000     -H(W1,S24)     H(S13)     H(W2)>=0
005-th inequality: weight = 1.000000     -H(S13)     H(W1,S31)     H(S13,S24)    -H(W1,S23,S41)>=0
006-th inequality: weight = 1.000000     -H(W1,W2,W3,W4)    -H(W1,S31)     H(W1,W4,S21)     H(W1,S24,S41)>=0
007-th inequality: weight = 1.000000     -H(W1,W2,W3,W4)     H(W1,W3,S21,S41)     H(W1,W3,S21,S23,S41)    -H(W1,S24,S31,S41)>=0
008-th inequality: weight = 1.000000     -H(W1,W2,W3,W4)     H(W1,S24)    -H(S13,S24)     H(W1,S23,S41)>=0
009-th inequality: weight = 4.000000     -H(W2)       A>=0
010-th inequality: weight = 6.000000     -H(S13)       B>=0
011-th inequality: weight = 3.000000      H(W1,W2,W3,W4)    -1.0*>=0
******************************************************************
\end{verbatim}
}

\end{center}
The result can be interpreted as follows. The bounds
\begin{align*}
 2A+B\geq 1,\quad 4A+6B\geq 3
\end{align*}
in the problem description file can be proved by adding the 5 inequalities and the 11 inequalities shown, with the weights given for each one. A constant value is marked using the ``*''. Note that the proof is solved twice, one using float point values, and the other as an integer program.

The last function is to performance sensitivity analysis
\begin{verbatim}
       CplexCompute PDRG4x3x3.pd sensitivity
\end{verbatim}
and we obtain the following result 
\begin{center} 
{\tt \scriptsize
\begin{verbatim}
Total number of elements before reduction: 65536
Total number of elements after reduction: 179
Total number of constraints given to Cplex: 40862
******************************************************************
Optimal value for A + B = 0.625000.
Sensitivity results:
Sensitivity A                        = [0.37500, 0.37500]
Sensitivity B                        = [0.25000, 0.25000]
Sensitivity 2I(S12;S21|S32) + H(S21|S31) + A= [0.87500, 0.87500]
******************************************************************
\end{verbatim}
}
\end{center}
In this case, there does not exist any slack at this optimal value in these quantities. 

\section{Errors, Warnings, and Auxiliary Functions}

\subsection{Errors and Warnings}

Most of the error or warning messages will be encountered due to problem description syntax errors. For example, when the same random variable appears in the same row of the permutation twice, or a given line has a character not allowed in that section, e.g., ``$|$'' in the dependency section. 

One particular error checking we perform is to make sure that the input permutations form a valid permutation group. If they do not form a permutation group, then the reduction used in the program may not be valid, and as a consequence, the computed result may not be a valid bound. This type of error may trigger a message such as
\begin{verbatim}
  Bad Symmetry -- missing permutation ...
\end{verbatim}
If this message appears, the reader is advised to carefully examine the permutation section to eliminate any input error. 

\subsection{Auxiliary Functions on Symmetry Relations}

The symmetry structure permutations become tedious to write manually for complex problems. One can use any scripting language, such as Matlab or Python, to produce the this section of the problem description file. Python scripts are included in the package for several well known coding problems.

\section{A Remark on Problem Formulation}

Despite the various reductions we employ in the toolbox, the computation scale is still usually quite large. As a rule of thumb, the toolbox can usually handle problems with less than equal to 20 random variables handily, which in many cases has already proved to provide  surprisingly powerful insights \cite{Tian:JSAC13,TianLiu:15,Tian:16Computer}. Depending on the problem structure, particularly the symmetry structure, it might be possible to produce meaningful results for problems with 25$\sim$27 random variables. Problems at an even large scale are likely to trigger memory issue or cause the overflow due to the word length restriction. 

Problem formulation plays an extremely important role in utilizing this toolbox. Let us reconsider the regenerating code problem with parameter $(n,k,d)=(5,4,4)$. A simple extension of the $(4,3,3)$ formulation we discuss above will lead to a total of $25$ random variables in the problem, which, though possible, poses a challenge in both memory usage and computation time. However, notice that we can in fact eliminate the random variables $(W_1,W_2,\ldots,W_5)$ altogether, by simply viewing $W_i$ as the collection of $\{(S_{i,j},j\neq i\}$. This helps to reduce the number of variables to $20$, and the computation can be done without much difficulty in the toolbox. The first five sections of the problem description file are as follows in this case, and the we omitted the symmetry portion here for better formatting.

{\tt \tiny
\begin{verbatim}
PD
{
  "RV": ["S12","S13","S14","S15","S21","S23","S24","S25","S31","S32","S34","S35","S41","S42","S43","S45","S51","S52","S53","S54"],
  "AL": ["A","B"],
  "O": "3.2A + 2B",
  "D": [
    {"dependent":["S12","S13","S14","S15"],"given":["S21","S31","S41","S51"]},
    {"dependent":["S21","S23","S24","S25"],"given":["S12","S32","S42","S52"]},
    {"dependent":["S31","S32","S34","S35"],"given":["S13","S23","S43","S53"]},
    {"dependent":["S41","S42","S43","S45"],"given":["S14","S24","S34","S54"]},
    {"dependent":["S51","S52","S53","S54"],"given":["S15","S25","S35","S45"]} ],
  "I": [],
  "BC": [
    "H(S12,S13,S14,S15) - A <= 0",
    "H(S12) - B <= 0",
    "H(S12,S13,S14,S15,S21,S23,S24,S25,S31,S32,S34,S35,S41,S42,S43,S45,S51,S52,S53,S54) >= 1" ],
  "BP": ["15A+10B >= 6"],
  "QU": ["A","B", "H(S13|S14,S24)","I(S13;S24|S25)+2I(S32;S32)"],  
  "S": [
    ["S12","S13","S14","S15","S21","S23","S24","S25","S31","S32","S34","S35","S41","S42","S43","S45","S51","S52","S53","S54"],
    ["S54","S53","S52","S51","S45","S43","S42","S41","S35","S34","S32","S31","S25","S24","S23","S21","S15","S14","S13","S12"],
    ["S54","S53","S51","S52","S45","S43","S41","S42","S35","S34","S31","S32","S15","S14","S13","S12","S25","S24","S23","S21"],
    ["S54","S52","S53","S51","S45","S42","S43","S41","S25","S24","S23","S21","S35","S34","S32","S31","S15","S14","S12","S13"],
    ["S54","S52","S51","S53","S45","S42","S41","S43","S25","S24","S21","S23","S15","S14","S12","S13","S35","S34","S32","S31"],
    ["S54","S51","S53","S52","S45","S41","S43","S42","S15","S14","S13","S12","S35","S34","S31","S32","S25","S24","S21","S23"],
    ["S54","S51","S52","S53","S45","S41","S42","S43","S15","S14","S12","S13","S25","S24","S21","S23","S35","S34","S31","S32"],
    ["S53","S54","S52","S51","S35","S34","S32","S31","S45","S43","S42","S41","S25","S23","S24","S21","S15","S13","S14","S12"],
    ["S53","S54","S51","S52","S35","S34","S31","S32","S45","S43","S41","S42","S15","S13","S14","S12","S25","S23","S24","S21"],
    ["S53","S52","S54","S51","S35","S32","S34","S31","S25","S23","S24","S21","S45","S43","S42","S41","S15","S13","S12","S14"],
    ["S53","S52","S51","S54","S35","S32","S31","S34","S25","S23","S21","S24","S15","S13","S12","S14","S45","S43","S42","S41"],
    ["S53","S51","S54","S52","S35","S31","S34","S32","S15","S13","S14","S12","S45","S43","S41","S42","S25","S23","S21","S24"],
    ["S53","S51","S52","S54","S35","S31","S32","S34","S15","S13","S12","S14","S25","S23","S21","S24","S45","S43","S41","S42"],
    ["S52","S54","S53","S51","S25","S24","S23","S21","S45","S42","S43","S41","S35","S32","S34","S31","S15","S12","S14","S13"],
    ["S52","S54","S51","S53","S25","S24","S21","S23","S45","S42","S41","S43","S15","S12","S14","S13","S35","S32","S34","S31"],
    ["S52","S53","S54","S51","S25","S23","S24","S21","S35","S32","S34","S31","S45","S42","S43","S41","S15","S12","S13","S14"],
    ["S52","S53","S51","S54","S25","S23","S21","S24","S35","S32","S31","S34","S15","S12","S13","S14","S45","S42","S43","S41"],
    ["S52","S51","S54","S53","S25","S21","S24","S23","S15","S12","S14","S13","S45","S42","S41","S43","S35","S32","S31","S34"],
    ["S52","S51","S53","S54","S25","S21","S23","S24","S15","S12","S13","S14","S35","S32","S31","S34","S45","S42","S41","S43"],
    ["S51","S54","S53","S52","S15","S14","S13","S12","S45","S41","S43","S42","S35","S31","S34","S32","S25","S21","S24","S23"],
    ["S51","S54","S52","S53","S15","S14","S12","S13","S45","S41","S42","S43","S25","S21","S24","S23","S35","S31","S34","S32"],
    ["S51","S53","S54","S52","S15","S13","S14","S12","S35","S31","S34","S32","S45","S41","S43","S42","S25","S21","S23","S24"],
    ["S51","S53","S52","S54","S15","S13","S12","S14","S35","S31","S32","S34","S25","S21","S23","S24","S45","S41","S43","S42"],
    ["S51","S52","S54","S53","S15","S12","S14","S13","S25","S21","S24","S23","S45","S41","S42","S43","S35","S31","S32","S34"],
    ["S51","S52","S53","S54","S15","S12","S13","S14","S25","S21","S23","S24","S35","S31","S32","S34","S45","S41","S42","S43"],
    ["S45","S43","S42","S41","S54","S53","S52","S51","S34","S35","S32","S31","S24","S25","S23","S21","S14","S15","S13","S12"],
    ["S45","S43","S41","S42","S54","S53","S51","S52","S34","S35","S31","S32","S14","S15","S13","S12","S24","S25","S23","S21"],
    ["S45","S42","S43","S41","S54","S52","S53","S51","S24","S25","S23","S21","S34","S35","S32","S31","S14","S15","S12","S13"],
    ["S45","S42","S41","S43","S54","S52","S51","S53","S24","S25","S21","S23","S14","S15","S12","S13","S34","S35","S32","S31"],
    ["S45","S41","S43","S42","S54","S51","S53","S52","S14","S15","S13","S12","S34","S35","S31","S32","S24","S25","S21","S23"],
    ["S45","S41","S42","S43","S54","S51","S52","S53","S14","S15","S12","S13","S24","S25","S21","S23","S34","S35","S31","S32"],
    ["S43","S45","S42","S41","S34","S35","S32","S31","S54","S53","S52","S51","S24","S23","S25","S21","S14","S13","S15","S12"],
    ["S43","S45","S41","S42","S34","S35","S31","S32","S54","S53","S51","S52","S14","S13","S15","S12","S24","S23","S25","S21"],
    ["S43","S42","S45","S41","S34","S32","S35","S31","S24","S23","S25","S21","S54","S53","S52","S51","S14","S13","S12","S15"],
    ["S43","S42","S41","S45","S34","S32","S31","S35","S24","S23","S21","S25","S14","S13","S12","S15","S54","S53","S52","S51"],
    ["S43","S41","S45","S42","S34","S31","S35","S32","S14","S13","S15","S12","S54","S53","S51","S52","S24","S23","S21","S25"],
    ["S43","S41","S42","S45","S34","S31","S32","S35","S14","S13","S12","S15","S24","S23","S21","S25","S54","S53","S51","S52"],
    ["S42","S45","S43","S41","S24","S25","S23","S21","S54","S52","S53","S51","S34","S32","S35","S31","S14","S12","S15","S13"],
    ["S42","S45","S41","S43","S24","S25","S21","S23","S54","S52","S51","S53","S14","S12","S15","S13","S34","S32","S35","S31"],
    ["S42","S43","S45","S41","S24","S23","S25","S21","S34","S32","S35","S31","S54","S52","S53","S51","S14","S12","S13","S15"],
    ["S42","S43","S41","S45","S24","S23","S21","S25","S34","S32","S31","S35","S14","S12","S13","S15","S54","S52","S53","S51"],
    ["S42","S41","S45","S43","S24","S21","S25","S23","S14","S12","S15","S13","S54","S52","S51","S53","S34","S32","S31","S35"],
    ["S42","S41","S43","S45","S24","S21","S23","S25","S14","S12","S13","S15","S34","S32","S31","S35","S54","S52","S51","S53"],
    ["S41","S45","S43","S42","S14","S15","S13","S12","S54","S51","S53","S52","S34","S31","S35","S32","S24","S21","S25","S23"],
    ["S41","S45","S42","S43","S14","S15","S12","S13","S54","S51","S52","S53","S24","S21","S25","S23","S34","S31","S35","S32"],
    ["S41","S43","S45","S42","S14","S13","S15","S12","S34","S31","S35","S32","S54","S51","S53","S52","S24","S21","S23","S25"],
    ["S41","S43","S42","S45","S14","S13","S12","S15","S34","S31","S32","S35","S24","S21","S23","S25","S54","S51","S53","S52"],
    ["S41","S42","S45","S43","S14","S12","S15","S13","S24","S21","S25","S23","S54","S51","S52","S53","S34","S31","S32","S35"],
    ["S41","S42","S43","S45","S14","S12","S13","S15","S24","S21","S23","S25","S34","S31","S32","S35","S54","S51","S52","S53"],
    ["S35","S34","S32","S31","S53","S54","S52","S51","S43","S45","S42","S41","S23","S25","S24","S21","S13","S15","S14","S12"],
    ["S35","S34","S31","S32","S53","S54","S51","S52","S43","S45","S41","S42","S13","S15","S14","S12","S23","S25","S24","S21"],
    ["S35","S32","S34","S31","S53","S52","S54","S51","S23","S25","S24","S21","S43","S45","S42","S41","S13","S15","S12","S14"],
    ["S35","S32","S31","S34","S53","S52","S51","S54","S23","S25","S21","S24","S13","S15","S12","S14","S43","S45","S42","S41"],
    ["S35","S31","S34","S32","S53","S51","S54","S52","S13","S15","S14","S12","S43","S45","S41","S42","S23","S25","S21","S24"],
    ["S35","S31","S32","S34","S53","S51","S52","S54","S13","S15","S12","S14","S23","S25","S21","S24","S43","S45","S41","S42"],
    ["S34","S35","S32","S31","S43","S45","S42","S41","S53","S54","S52","S51","S23","S24","S25","S21","S13","S14","S15","S12"],
    ["S34","S35","S31","S32","S43","S45","S41","S42","S53","S54","S51","S52","S13","S14","S15","S12","S23","S24","S25","S21"],
    ["S34","S32","S35","S31","S43","S42","S45","S41","S23","S24","S25","S21","S53","S54","S52","S51","S13","S14","S12","S15"],
    ["S34","S32","S31","S35","S43","S42","S41","S45","S23","S24","S21","S25","S13","S14","S12","S15","S53","S54","S52","S51"],
    ["S34","S31","S35","S32","S43","S41","S45","S42","S13","S14","S15","S12","S53","S54","S51","S52","S23","S24","S21","S25"],
    ["S34","S31","S32","S35","S43","S41","S42","S45","S13","S14","S12","S15","S23","S24","S21","S25","S53","S54","S51","S52"],
    ["S32","S35","S34","S31","S23","S25","S24","S21","S53","S52","S54","S51","S43","S42","S45","S41","S13","S12","S15","S14"],
    ["S32","S35","S31","S34","S23","S25","S21","S24","S53","S52","S51","S54","S13","S12","S15","S14","S43","S42","S45","S41"],
    ["S32","S34","S35","S31","S23","S24","S25","S21","S43","S42","S45","S41","S53","S52","S54","S51","S13","S12","S14","S15"],
    ["S32","S34","S31","S35","S23","S24","S21","S25","S43","S42","S41","S45","S13","S12","S14","S15","S53","S52","S54","S51"],
    ["S32","S31","S35","S34","S23","S21","S25","S24","S13","S12","S15","S14","S53","S52","S51","S54","S43","S42","S41","S45"],
    ["S32","S31","S34","S35","S23","S21","S24","S25","S13","S12","S14","S15","S43","S42","S41","S45","S53","S52","S51","S54"],
    ["S31","S35","S34","S32","S13","S15","S14","S12","S53","S51","S54","S52","S43","S41","S45","S42","S23","S21","S25","S24"],
    ["S31","S35","S32","S34","S13","S15","S12","S14","S53","S51","S52","S54","S23","S21","S25","S24","S43","S41","S45","S42"],
    ["S31","S34","S35","S32","S13","S14","S15","S12","S43","S41","S45","S42","S53","S51","S54","S52","S23","S21","S24","S25"],
    ["S31","S34","S32","S35","S13","S14","S12","S15","S43","S41","S42","S45","S23","S21","S24","S25","S53","S51","S54","S52"],
    ["S31","S32","S35","S34","S13","S12","S15","S14","S23","S21","S25","S24","S53","S51","S52","S54","S43","S41","S42","S45"],
    ["S31","S32","S34","S35","S13","S12","S14","S15","S23","S21","S24","S25","S43","S41","S42","S45","S53","S51","S52","S54"],
    ["S25","S24","S23","S21","S52","S54","S53","S51","S42","S45","S43","S41","S32","S35","S34","S31","S12","S15","S14","S13"],
    ["S25","S24","S21","S23","S52","S54","S51","S53","S42","S45","S41","S43","S12","S15","S14","S13","S32","S35","S34","S31"],
    ["S25","S23","S24","S21","S52","S53","S54","S51","S32","S35","S34","S31","S42","S45","S43","S41","S12","S15","S13","S14"],
    ["S25","S23","S21","S24","S52","S53","S51","S54","S32","S35","S31","S34","S12","S15","S13","S14","S42","S45","S43","S41"],
    ["S25","S21","S24","S23","S52","S51","S54","S53","S12","S15","S14","S13","S42","S45","S41","S43","S32","S35","S31","S34"],
    ["S25","S21","S23","S24","S52","S51","S53","S54","S12","S15","S13","S14","S32","S35","S31","S34","S42","S45","S41","S43"],
    ["S24","S25","S23","S21","S42","S45","S43","S41","S52","S54","S53","S51","S32","S34","S35","S31","S12","S14","S15","S13"],
    ["S24","S25","S21","S23","S42","S45","S41","S43","S52","S54","S51","S53","S12","S14","S15","S13","S32","S34","S35","S31"],
    ["S24","S23","S25","S21","S42","S43","S45","S41","S32","S34","S35","S31","S52","S54","S53","S51","S12","S14","S13","S15"],
    ["S24","S23","S21","S25","S42","S43","S41","S45","S32","S34","S31","S35","S12","S14","S13","S15","S52","S54","S53","S51"],
    ["S24","S21","S25","S23","S42","S41","S45","S43","S12","S14","S15","S13","S52","S54","S51","S53","S32","S34","S31","S35"],
    ["S24","S21","S23","S25","S42","S41","S43","S45","S12","S14","S13","S15","S32","S34","S31","S35","S52","S54","S51","S53"],
    ["S23","S25","S24","S21","S32","S35","S34","S31","S52","S53","S54","S51","S42","S43","S45","S41","S12","S13","S15","S14"],
    ["S23","S25","S21","S24","S32","S35","S31","S34","S52","S53","S51","S54","S12","S13","S15","S14","S42","S43","S45","S41"],
    ["S23","S24","S25","S21","S32","S34","S35","S31","S42","S43","S45","S41","S52","S53","S54","S51","S12","S13","S14","S15"],
    ["S23","S24","S21","S25","S32","S34","S31","S35","S42","S43","S41","S45","S12","S13","S14","S15","S52","S53","S54","S51"],
    ["S23","S21","S25","S24","S32","S31","S35","S34","S12","S13","S15","S14","S52","S53","S51","S54","S42","S43","S41","S45"],
    ["S23","S21","S24","S25","S32","S31","S34","S35","S12","S13","S14","S15","S42","S43","S41","S45","S52","S53","S51","S54"],
    ["S21","S25","S24","S23","S12","S15","S14","S13","S52","S51","S54","S53","S42","S41","S45","S43","S32","S31","S35","S34"],
    ["S21","S25","S23","S24","S12","S15","S13","S14","S52","S51","S53","S54","S32","S31","S35","S34","S42","S41","S45","S43"],
    ["S21","S24","S25","S23","S12","S14","S15","S13","S42","S41","S45","S43","S52","S51","S54","S53","S32","S31","S34","S35"],
    ["S21","S24","S23","S25","S12","S14","S13","S15","S42","S41","S43","S45","S32","S31","S34","S35","S52","S51","S54","S53"],
    ["S21","S23","S25","S24","S12","S13","S15","S14","S32","S31","S35","S34","S52","S51","S53","S54","S42","S41","S43","S45"],
    ["S21","S23","S24","S25","S12","S13","S14","S15","S32","S31","S34","S35","S42","S41","S43","S45","S52","S51","S53","S54"],
    ["S15","S14","S13","S12","S51","S54","S53","S52","S41","S45","S43","S42","S31","S35","S34","S32","S21","S25","S24","S23"],
    ["S15","S14","S12","S13","S51","S54","S52","S53","S41","S45","S42","S43","S21","S25","S24","S23","S31","S35","S34","S32"],
    ["S15","S13","S14","S12","S51","S53","S54","S52","S31","S35","S34","S32","S41","S45","S43","S42","S21","S25","S23","S24"],
    ["S15","S13","S12","S14","S51","S53","S52","S54","S31","S35","S32","S34","S21","S25","S23","S24","S41","S45","S43","S42"],
    ["S15","S12","S14","S13","S51","S52","S54","S53","S21","S25","S24","S23","S41","S45","S42","S43","S31","S35","S32","S34"],
    ["S15","S12","S13","S14","S51","S52","S53","S54","S21","S25","S23","S24","S31","S35","S32","S34","S41","S45","S42","S43"],
    ["S14","S15","S13","S12","S41","S45","S43","S42","S51","S54","S53","S52","S31","S34","S35","S32","S21","S24","S25","S23"],
    ["S14","S15","S12","S13","S41","S45","S42","S43","S51","S54","S52","S53","S21","S24","S25","S23","S31","S34","S35","S32"],
    ["S14","S13","S15","S12","S41","S43","S45","S42","S31","S34","S35","S32","S51","S54","S53","S52","S21","S24","S23","S25"],
    ["S14","S13","S12","S15","S41","S43","S42","S45","S31","S34","S32","S35","S21","S24","S23","S25","S51","S54","S53","S52"],
    ["S14","S12","S15","S13","S41","S42","S45","S43","S21","S24","S25","S23","S51","S54","S52","S53","S31","S34","S32","S35"],
    ["S14","S12","S13","S15","S41","S42","S43","S45","S21","S24","S23","S25","S31","S34","S32","S35","S51","S54","S52","S53"],
    ["S13","S15","S14","S12","S31","S35","S34","S32","S51","S53","S54","S52","S41","S43","S45","S42","S21","S23","S25","S24"],
    ["S13","S15","S12","S14","S31","S35","S32","S34","S51","S53","S52","S54","S21","S23","S25","S24","S41","S43","S45","S42"],
    ["S13","S14","S15","S12","S31","S34","S35","S32","S41","S43","S45","S42","S51","S53","S54","S52","S21","S23","S24","S25"],
    ["S13","S14","S12","S15","S31","S34","S32","S35","S41","S43","S42","S45","S21","S23","S24","S25","S51","S53","S54","S52"],
    ["S13","S12","S15","S14","S31","S32","S35","S34","S21","S23","S25","S24","S51","S53","S52","S54","S41","S43","S42","S45"],
    ["S13","S12","S14","S15","S31","S32","S34","S35","S21","S23","S24","S25","S41","S43","S42","S45","S51","S53","S52","S54"],
    ["S12","S15","S14","S13","S21","S25","S24","S23","S51","S52","S54","S53","S41","S42","S45","S43","S31","S32","S35","S34"],
    ["S12","S15","S13","S14","S21","S25","S23","S24","S51","S52","S53","S54","S31","S32","S35","S34","S41","S42","S45","S43"],
    ["S12","S14","S15","S13","S21","S24","S25","S23","S41","S42","S45","S43","S51","S52","S54","S53","S31","S32","S34","S35"],
    ["S12","S14","S13","S15","S21","S24","S23","S25","S41","S42","S43","S45","S31","S32","S34","S35","S51","S52","S54","S53"],
    ["S12","S13","S15","S14","S21","S23","S25","S24","S31","S32","S35","S34","S51","S52","S53","S54","S41","S42","S43","S45"] ]
}

\end{verbatim}
}

With this representation, the computation for the simple bounding will then produce results in about 21 seconds. 

\section{Conclusion and Future Work}

We provide an open source toolbox for computer-aided investigation of the fundamental limits of information systems. The given program is designed to read a problem description file, convert it to the corresponding LP, and then produce meaningful bounds directly without much user intervention.

\section*{Acknowledgment}

This work is supported in part by the National Science Foundation through grants CCF-18-16518 and CCF-18-16546. 

\bibliographystyle{IEEEtran}

\begin{thebibliography}{10}
\providecommand{\url}[1]{#1}
\csname url@samestyle\endcsname
\providecommand{\newblock}{\relax}
\providecommand{\bibinfo}[2]{#2}
\providecommand{\BIBentrySTDinterwordspacing}{\spaceskip=0pt\relax}
\providecommand{\BIBentryALTinterwordstretchfactor}{4}
\providecommand{\BIBentryALTinterwordspacing}{\spaceskip=\fontdimen2\font plus
\BIBentryALTinterwordstretchfactor\fontdimen3\font minus
  \fontdimen4\font\relax}
\providecommand{\BIBforeignlanguage}[2]{{%
\expandafter\ifx\csname l@#1\endcsname\relax
\typeout{** WARNING: IEEEtran.bst: No hyphenation pattern has been}%
\typeout{** loaded for the language `#1'. Using the pattern for}%
\typeout{** the default language instead.}%
\else
\language=\csname l@#1\endcsname
\fi
#2}}
\providecommand{\BIBdecl}{\relax}
\BIBdecl

\bibitem{Yeung:97}
R.~W. Yeung, ``A framework for linear information inequalities,'' \emph{IEEE
  Transactions on Information Theory}, vol.~43, no.~6, pp. 1924--1934, Nov.
  1997.

\bibitem{ITIP}
\BIBentryALTinterwordspacing
\emph{{Information Theoretic Inequality Prover (ITIP)}}, 1998-2008. [Online].
  Available: \url{http://user-www.ie.cuhk.edu.hk/~ITIP/}
\BIBentrySTDinterwordspacing

\bibitem{XITIP}
\BIBentryALTinterwordspacing
\emph{{Xitip: Information Theoretic Inequalities Prover}}, 2007. [Online].
  Available: \url{http://xitip.epfl.ch/}
\BIBentrySTDinterwordspacing

\bibitem{yeung2002information}
R.~W. Yeung, T.~T. Lee, and Z.~Ye, ``Information-theoretic characterizations of
  conditional mutual independence and markov random fields,'' \emph{IEEE
  Transactions on Information Theory}, vol.~48, no.~7, pp. 1996--2011, 2002.

\bibitem{dougherty2006six}
R.~Dougherty, C.~Freiling, and K.~Zeger, ``Six new non-shannon information
  inequalities,'' in \emph{2006 IEEE International Symposium on Information
  Theory}.\hskip 1em plus 0.5em minus 0.4em\relax IEEE, 2006, pp. 233--236.

\bibitem{dougherty2011non}
------, ``Non-shannon information inequalities in four random variables,''
  \emph{arXiv preprint arXiv:1104.3602}, 2011.

\bibitem{Tian:JSAC13}
C.~Tian, ``Characterizing the rate region of the (4, 3, 3) exact-repair
  regenerating codes,'' \emph{IEEE Journal on Selected Areas in
  Communications}, vol.~32, no.~5, pp. 967--975, May 2014.

\bibitem{TianLiu:15}
C.~Tian and T.~Liu, ``Multilevel diversity coding with regeneration,''
  \emph{IEEE Transactions on Information Theory}, vol.~62, no.~9, pp.
  4833--4847, Sep. 2016.

\bibitem{Tian:16Computer}
C.~Tian, ``Symmetry, outer bounds, and code constructions: A computer-aided
  investigation on the fundamental limits of caching,'' \emph{Entropy},
  vol.~20, no.~8, pp. 603.1--43, Aug. 2018.

\bibitem{tian2018caching}
C.~Tian and J.~Chen, ``Caching and delivery via interference elimination,''
  \emph{IEEE Transactions on Information Theory}, vol.~64, no.~3, pp.
  1548--1560, 2018.

\bibitem{Tian_Sun_Chen_Storage}
C.~Tian, H.~Sun, and J.~Chen, ``A {Shannon}-theoretic approach to the
  storage-retrieval tradeoff in {PIR} systems,'' in \emph{2018 Proceedings of
  IEEE International Symposium on Information Theory (ISIT)}, Jun. 2018, pp.
  1904--1908.

\bibitem{tian2020storage}
C.~Tian, ``On the storage cost of private information retrieval,'' \emph{IEEE
  Transactions on Information Theory}, 2020.

\bibitem{li2016multilevel}
C.~Li, S.~Weber, and J.~M. Walsh, ``Multilevel diversity coding systems: Rate
  regions, codes, computation, \& forbidden minors,'' \emph{IEEE Transactions
  on Information Theory}, vol.~63, no.~1, pp. 230--251, 2016.

\bibitem{ho2020proving}
S.-W. Ho, L.~Ling, C.~W. Tan, and R.~W. Yeung, ``Proving and disproving
  information inequalities: Theory and scalable algorithms,'' \emph{IEEE
  Transactions on Information Theory}, 2020.

\bibitem{chan2019minimal}
T.~Chan, S.~Thakor, and A.~Grant, ``Minimal characterization of shannon-type
  inequalities under functional dependence and full conditional independence
  structures,'' \emph{IEEE Transactions on Information Theory}, vol.~65, no.~7,
  pp. 4041--4051, 2019.

\bibitem{ZhangTian:17TCOM}
K.~Zhang and C.~Tian, ``On the symmetry reduction of information
  inequalities,'' \emph{IEEE Transactions on Communications}, vol.~66, no.~6,
  pp. 2396--2408, 2018.

\bibitem{Walsh:16}
C.~Li, S.~Weber, and J.~M. Walsh, ``Multilevel diversity coding systems: Rate
  regions, codes, computation, \& forbidden minors,'' \emph{IEEE Transactions
  on Information Theory}, vol.~63, no.~1, pp. 230--251, Jan. 2017.

\bibitem{li2017multi}
------, ``On multi-source networks: Enumeration, rate region computation, and
  hierarchy,'' \emph{IEEE Transactions on Information Theory}, vol.~63, no.~11,
  pp. 7283--7303, 2017.

\bibitem{apte2015exploiting}
J.~Apte and J.~M. Walsh, ``Exploiting symmetry in computing polyhedral bounds
  on network coding rate regions,'' in \emph{2015 International symposium on
  network coding (NetCod)}, 2015, pp. 76--80.

\bibitem{li2013new}
C.~Li, J.~Apte, J.~M. Walsh, and S.~Weber, ``A new computational approach for
  determining rate regions and optimal codes for coded networks,'' in
  \emph{2013 International Symposium on Network Coding (NetCod)}, 2013, pp.
  1--6.

\bibitem{apte2014algorithms}
J.~Apte, C.~Li, and J.~M. Walsh, ``Algorithms for computing network coding rate
  regions via single element extensions of matroids,'' in \emph{2014 IEEE
  International Symposium on Information Theory}, 2014, pp. 2306--2310.

\bibitem{Permuter:15newsletter}
I.~B. Gattegno, Z.~Goldfeld, and H.~H. Permuter, ``Fourier-motzkin elimination
  software for information theoretic inequalities,'' \emph{IEEE Information
  Theory Newsletter}, vol.~65, no.~3, pp. 25--29, Sep. 2015.

\bibitem{gurpinar2019use}
E.~G{\"u}rp{\i}nar and A.~Romashchenko, ``How to use undiscovered information
  inequalities: Direct applications of the copy lemma,'' in \emph{2019 IEEE
  International Symposium on Information Theory (ISIT)}, 2019, pp. 1377--1381.

\bibitem{CoverThomas}
T.~M. Cover and J.~A. Thomas, \emph{Elements of Information Theory},
  1st~ed.\hskip 1em plus 0.5em minus 0.4em\relax New York: Wiley, 1991.

\bibitem{Yeung:book}
R.~Yeung, \emph{A First Course in Information Theory}.\hskip 1em plus 0.5em
  minus 0.4em\relax New York: Kluwer Academic Publishers, 2002.

\bibitem{yeung2008information}
R.~W. Yeung, \emph{Information theory and network coding}.\hskip 1em plus 0.5em
  minus 0.4em\relax Springer Science \& Business Media, 2008.

\bibitem{Zhang:97}
Z.~Zhang and R.~W. Yeung, ``A non-{Shannon}-type conditional inequality of
  information quantities,'' \emph{IEEE Transactions on Information Theory},
  vol.~43, no.~6, pp. 1982--1986, Nov. 1997.

\bibitem{matus2007infinitely}
F.~Matus, ``Infinitely many information inequalities,'' in \emph{2007 IEEE
  International Symposium on Information Theory}.\hskip 1em plus 0.5em minus
  0.4em\relax IEEE, 2007, pp. 41--44.

\bibitem{Dougherty:05}
R.~Dougherty, C.~Freiling, and K.~Zeger, ``Insufficiency of linear coding in
  network information flow,'' \emph{IEEE Transactions on Information Theory},
  vol.~51, no.~8, pp. 2745--2759, Aug. 2005.

\bibitem{dougherty2007networks}
------, ``Networks, matroids, and non-shannon information inequalities,''
  \emph{IEEE Transactions on Information Theory}, vol.~53, no.~6, pp.
  1949--1969, 2007.

\bibitem{dgw:10:nc}
A.~G. Dimakis, P.~B. Godfrey, Y.~Wu, M.~Wainwright, and K.~Ramchandran,
  ``Network coding for distributed storage systems,'' \emph{IEEE Transactions
  on Information Theory}, 2010.

\bibitem{drw:11:snc}
A.~G. Dimakis, K.~Ramchandran, Y.~Wu, and C.~Suh, ``A survey on network codes
  for distributed storage,'' \emph{Proceedings of the IEEE}, vol.~99, no.~3,
  March 2011.

\bibitem{MaddahAliNiesen:14}
M.~A. Maddah-Ali and U.~Niesen, ``Fundamental limits of caching,'' \emph{IEEE
  Transactions on Information Theory}, vol.~60, no.~5, pp. 2856--2867, May
  2014.

\bibitem{Lassez:92}
C.~Lassez and J.-L. Lassez, ``Quantifier elimination for conjunctions of linear
  constraints via a convex hull algorithm,'' in \emph{Symbolic and numerical
  computation for artificial intelligence}, B.~R. Donald, D.~Kapur, and J.~L.
  Mundy, Eds.\hskip 1em plus 0.5em minus 0.4em\relax San Diego, CA: Academic
  Press, 1992, ch.~4, pp. 103--1199.

\bibitem{ho2014proving}
S.-W. Ho, C.~W. Tan, and R.~W. Yeung, ``Proving and disproving information
  inequalities,'' in \emph{2014 IEEE International Symposium on Information
  Theory}.\hskip 1em plus 0.5em minus 0.4em\relax IEEE, 2014, pp. 2814--2818.

\bibitem{Cplex}
\BIBentryALTinterwordspacing
\emph{{IBM ILOG CPLEX Optimizer}}, IBM, 2009-2019. [Online]. Available:
  \url{http://www-01.ibm.com/software/integration/optimization/cplex-optimizer/}
\BIBentrySTDinterwordspacing

\bibitem{Gurobi}
\BIBentryALTinterwordspacing
\emph{{Gurobi Optimizer}}, Gurobi, 2008-2019. [Online]. Available:
  \url{http:/www.gurobi.com/}
\BIBentrySTDinterwordspacing

\end{thebibliography}
 \newcommand{\noop}[1]{}

\end{document}